\documentclass[pra,preprint,showpacs]{revtex4}
\usepackage[centertags]{amsmath}
\usepackage{amsfonts}
\usepackage{amssymb}
\usepackage{amsthm} 
\usepackage{newlfont}
\usepackage{epsfig}
\usepackage{amscd}
\usepackage{graphicx}
\usepackage{epsfig}
\usepackage{color}
\usepackage{amscd}

\newcommand{\beq}{\begin{equation}}
\newcommand{\eeq}{\end{equation}}
\newcommand{\ba}{\begin{array}}
\newcommand{\ea}{\end{array}}
\newcommand{\bea}{\begin{eqnarray}}
\newcommand{\eea}{\end{eqnarray}}

\begin{document}

\begin{center}
{\large \sc \bf  {{}
Remote  creation of a one-qubit mixed state through a short homogeneous spin-1/2 chain.}}

\vskip 15pt

{\large 
 A.I.~Zenchuk 
}

\vskip 8pt

{\it Institute of Problems of Chemical Physics, RAS,
Chernogolovka, Moscow reg., 142432, Russia, e-mail:   zenchuk@itp.ac.ru
 } 
\end{center}


\begin{abstract}
We consider a method of remote mixed state creation  of a one-qubit subsystem (receiver) in a spin-1/2 chain 
governed by the nearest-neighbor $XY$-Hamiltonian. 
Owing to the evolution of the chain along  with the variable local unitary 
transformation of the one- or two-qubit sender,
a large  variety  of   receiver states   can be created during some time interval 
{{{}}  starting with a fixed initial state of the whole quantum system}.  {{} These
states form the creatable region of the receiver's state-space.}
It is remarkable that,  having the two-qubit sender, 
a large  creatable region may be covered at a properly fixed time instant  $t_0$
  using just the variable 
local unitary transformation of the sender. In this case we have completely local  
control of the remote state creation. {{{}}   In general, for a given initial state, 
there are such receiver's states that  may not be  created using the above tool. {{}
These states form 
the unavailable region. In turn, this unavailable region might be the creatable region of 
another sender. Thus, in future,  we have a way to  share the whole receiver's state-space 
among the creatable regions of several senders.}}
 The effectiveness of  remote state creation is
characterized by the  density function of the creatable region. 

\end{abstract}

\maketitle

\section{Introduction}
\label{Sec:Introduction}

{{{}}
Remote state creation means the creation of 
 a needed state of some selected subsystem of a quantum system (receiver) using the local operations on 
 another subsystem (sender). 
First, this problem appeared as a teleportation problem of 
unknown state from the sender (Alica)
to the receiver (Bob) \cite{BBCJPW,BPMEWZ,BBMHP} using  pairs of 
entangled qubits \cite{YS1,YS2,ZZHE}. It is important to note that all  existing quantum teleportation algorithms 
use a classical channel of information transfer as a necessary constituent. 
Decreasing the necessary amount of classically transmitted information is one of the
directions of  development 
of  remote state preparation algorithms 
\cite{BDSSBW,BHLSW,PBGWK2,PBGWK,DLMRKBPVZBW,XLYG,G}.
Regarding  experimental realizations of remote state preparation, 
one should note the experiments with pairs of entangled photons, which are 
widely used for this purpose
\cite{ZZHE,BPMEWZ,BBMHP,PBGWK2,PBGWK,DLMRKBPVZBW}. 
The remote  preparation of a single-qubit state with all three controllable 
parameters was studied in refs.
\cite{PBGWK2,XLYG,PBGWK}. Emphasize that an inherent aspect is 
the entanglement between (some of) the qubits of 
sender and receiver in (almost) all the above references. In addition, the discord as a resource for  
remote state preparation was studied, for instance, in 
\cite{DLMRKBPVZBW,G}.
}

{{{}}
As a special case of the remote state creation, we point out 
the problem of pure one-qubit quntum state transfer in spin-1/2 chains.} This problem   was first formulated
in the well-known paper by Bose \cite{Bose} and now it represents a special area of
quantum information processing. 
Several methods of either perfect \cite{CDEL,ACDE,KS} or high-fidelity (probability)
\cite{KF,KZ,GKMT,FZ} state transfer
have been proposed and studied. 
Perhaps the best known systems are the spin chain
with properly adjusted 
coupling constants  (the so-called fully engineered spin chain) \cite{CDEL,ACDE,KS}
and  the  homogeneous spin chain with remote end nodes 
(the so-called boundary-controlled spin chain) \cite{GKMT,WLKGGB}.

It was noted that high-fidelity  state transfer requires the very rigorous adjustment of  the parameters of a 
chain such as the coupling constants \cite{CDEL,ACDE,KS} and/or the local 
distribution of  external 
magnetic field \cite{DZ}.
Such a chain is very sensitive to  perturbations of its parameters,  which lead, 
in particular, to significant decrease
of the state transfer fidelity   \cite{CRMF,ZASO,ZASO2,ZASO3,SAOZ}. Although the boundary-controlled chain
is much simpler to realize,
 the price for this is  the long  state transfer time, 
which significantly reduces the effectiveness of such a chain \cite{FKZ}.

Alternatively, the  transfer of  complete information about the initial 
mixed state of a given subsystem (sender)
 to another subsystem (receiver)
was proposed as a development  of the state transfer methods  \cite{Z_2012} 
(state-information transfer). Information transfer is not sensitive 
to  the  parameters of the transfer line \cite{Z_2012}.
After the information about the  sender's  state is obtained by  the receiver at some time instant,
the initial state of the sender may be recovered  (if needed) using the local (non-unitary) transformation, 
namely, by 
solving the system of linear algebraic equations. 

{{{}} Note, that the state transfer   described in the above quoted references 
does not explicitly 
uses the 
concepts of quantum correlations between the sender and receiver, although  they are responsible for 
that process. The relation between   state transfer and entanglement 
(as a measure of quantum correlations) was studied,
for instance, 
in \cite{Bose,VGIZ,GKMT,GMT,BB,DFZ,DZ}. In addition, based on the result in \cite{Z_2012} concerning
  information transfer, the so-called 
informational correlation \cite{Z_inf} was  introduced, showing the sensitivity degree of the 
receiver's state to the  local unitary transformations of the sender. This measure seems to be 
more closely related 
to   remote state creation.  }

{{{}}
The remote state creation algorithm proposed in this paper 
 combines the ideas of  both pure state transfer 
\cite{Bose,CDEL,ACDE,KS,KF,KZ,GKMT,FZ,WLKGGB}
and   
mixed state-information transfer \cite{Z_2012}. More precisely, 
we study the  creation problem of possible receiver states  at some instant $t$ 
   starting with some initial state of the whole  system and using only 
   the initial local unitary transformation $U^A$ of the sender. Herewith, the evolution 
   of the whole system is governed by a certain Hamiltonian.
 This  gives us a tool for remote receiver state creation   {{{}}   
using  the parameters of the unitary transformation 
 $U^A$ and the time $t$ as control  parameters of the state creation process.} }

{{{}} 
We point out that the role of the 
classical channel of 
information transfer  is the basic difference between our algorithm and  the state 
creation algorithms  studied in refs.
\cite{BDSSBW,BHLSW,PBGWK2,PBGWK,DLMRKBPVZBW,XLYG,G}. Traditionally, the 
classical channel is used to transmit  (part of) the classical 
information about a quantum state, while we use this channel to transmit only the 
information about the time instant required to register the needed state. 
Moreover, in our case, the classical channel is needed only in the simplest cases  and may be disregarded in 
general, as  explained in 
 Sec.\ref{Section:stcr} and demonstrated with an example of Sec.\ref{Section:4node}. 
 In this case,  state creation is completely quantum.  

For the purpose of remote state creation, we use a particular quantum system, 
a spin-1/2 chain. At this stage, we do not study the effect of  quantum 
correlations (measured via either 
the entanglement \cite{Wootters,HW,P,AFOV,HHHH},  the  discord \cite{HV,OZ,Zurek}, or the
 informational correlation \cite{Z_inf}) 
on   state-creation processes, postponing this aspect for further study. 
}
 
 {{{}}  Hereafter, by the state of a particular subsystem $S$ of 
 a quantum system we mean the reduced density matrix (the marginal matrix)
 \begin{eqnarray}\label{margB}
 \rho^S= Tr_{rest} \rho,
 \end{eqnarray}
 where $\rho$ is the density matrix of the whole system and the trace is calculated with 
 respect to the rest of the
 quantum 
 system.
 This means that some additional projection procedure is required to extract the needed 
 state of  subsystem $B$. Experimental realization of this projection is not studied here. 
 Of course,  state (\ref{margB}) is achievable much more simpler than the product state 
 $\rho = \rho^{rest} \otimes \rho^B$, 
 which would be of more interest (the trace in eq.(\ref{margB})
 becomes trivial in that case). 
 But the requirement of getting a product state would lead to 
 additional severe relations among the parameters 
 of the sender's initial state and perhaps would require an  increase 
 in the dimensionality of the sender's Hilbert space. All this would
 complicate the calculations. Thus,  we choose  state (\ref{margB}) as a 
 simpler case of remote state creation, allowing us to study a set of features of 
 the state creation process.
 }
 
{{{}}   Our algorithm  relates  the particular initial sender's state 
with the proper receiver's state at some time instant. Therefore,
the considered process may be viewed as a 
map (not unique, in general) of the initially prepared sender's states to the proper receiver's state. 
Thus, 
keeping the term  ''state creation'', we  specify  the state-creation tool, 
which involves two initially controllable steps: (i) the creation of the selected  initial 
state of the whole system (ii) 
the implementation of the 
appropriate local unitary transformation  of the sender with the purpose of creating the needed receiver state. 
Therewith, the parameters of  unitary transformation are referred to as the  control parameters.
After these two initial steps,
the receiver's state is  ''built'' in an uncontrollable way through the transfer of
 quantum information about the sender's state \cite{Z_2012}. 
 This transfer is realized owing to the evolution of the 
 whole quantum system governed by a certain Hamiltonian.  To emphasize this feature, we call our
 creation process 
''remote state creation through  quantum information transfer''. 
}

 The problem of  remote state creation in a spin-1/2 chain using a sender and a receiver of different
 dimensionalities  $N^A$ and $N^B$ is a very complicated multi-parametric one. In this paper, 
 after representing some general statements regarding this process,  we 
 concentrate on the particular examples of short homogeneous chains with a one-qubit receiver 
 (the end-node of the chain)
 and a one(two)-qubit sender (the first node(s) of the chain). 
 We consider  state creation during a time interval $0\le t\le T$ with a fixed $T$ 
 {{{}}  (note that the parameter $T$
appears owing to the periodicity of the evolution  of the considered finite system;
this parameter depends on the 
 smallest (by absolute value) eigenvalue of the Hamiltonian)}
 and show that, if we use a two-qubit sender  ($N^A=4>N^B=2$),
 a large variety of  receiver states may be created  at some properly fixed 
 time instant $t_0$, $0\le t_0\le T$, using just the local  
 unitary transformation $U^A$ with variable parameters. 
 This effect is impossible in the case of a one-qubit sender, i.e., when 
 $N^A=N^B=2$. 
 
 {{{}}  
 We  point out the fact that, in general, there are  receiver states which  can not be created using 
 the above creation tool. These states form the unavailable region in the whole receiver state-space.
 This is an interesting characteristics of the state creation process. At the first glance, 
 it restricts the capability of the proposed state creation mechanism. However, this property might 
 allow us to divide the whole space of the receiver states into  subspaces controllable by different senders. 
 The use of such splitting is evident but we leave the problem  of sharing the receiver's state-space among 
 several senders beyond the scope of this  paper.}
 
 The paper is organized as follows.
  General ideas on the state creation as  a  map of 
 the control parameters of the sender to the required parameters of the receiver are formulated in
 Sec.\ref{Section:stcr} {{{}}  
 for an arbitrary quantum system governed by a certain Hamiltonian}. Mixed state creation in a homogeneous
 spin-1/2 chain governed by the nearest-neighbor 
 $XY$-Hamiltonian  with a  one-qubit receiver and one- or
 two-qubit sender is studied in Sec.\ref{Section:examples}. Basic results are briefly discussed in 
 Sec.\ref{Section:Conclusion}. Auxiliary information (and calculations) 
 regarding  one-qubit pure state transfer, some details on  state creation with one- and two-qubit sender, and 
 the basis of the Lie algebra of
 $SU(4)$ is presented in the Appendix, 
 Sec.\ref{Section:appendix}.

 \section{Remote state-creation of a subsystem of a quantum system.}
\label{Section:stcr}

{{{}}   \subsection{General state creation algorithm}}
{{{}}
In this section, we consider general aspects of the remote state creation through 
the quantum information transfer.}
To simplify our calculations, we  
deal with a particular type of the initial states, namely, the states 
representable by the tensor product of three diagonal
blocks:
\begin{eqnarray}\label{inst}
\rho_0=\rho^A_0\otimes \rho^C_0 \otimes \rho^B_0.
\end{eqnarray}
Here $\rho^A_0$, $\rho^B_0$ and $\rho^C_0$ describe the initial states of 
the sender,   receiver and  transmission line respectively. Being diagonal, these matrices are composed by the 
eigenvalues of the initial state. 
The remote state creation algorithm can be splitted into  the following steps. 
\begin{enumerate}
\item
Create the initial state of the sender, receiver and transmission line.
\item
Apply the unitary  transformation $U(\varphi)$ to the subsystem $A$ to obtain the  new  
initial density matrix $\rho_0(\varphi)$:
\begin{eqnarray}\label{instU}
\rho_0(\varphi)=(U^A(\varphi)\rho^A_0 (U^A(\varphi))^+)\otimes \rho^C_0 \otimes \rho^B_0,
\end{eqnarray}
where $\varphi=\{\varphi_1,\dots, \varphi_{(N^A)^2-1}\}$ 
is the list of parameters of the unitary transformation $U^A\in SU(N^A)$ 
which may vary in an arbitrary way. 
However, not all $(N^A)^2-1$ parameters of this transformation may affect the receiver's state 
(as is demonstrated below in this section) and, in principle, some additional constraints may be imposed on these
parameters.   
The choice of  parameters $\varphi_i$
is predicted by the needed receiver's state.
\item
Switch on the quantum evolution governed by a certain  Hamiltonian $H$
in accordance with the  Liouville equation
\begin{eqnarray}\label{ev}
\rho(\varphi,t)= e^{-i t {{H}}} \rho_0(\varphi) e^{i t {{H}}}.
\end{eqnarray}
{{{}}  The information about the initial sender's state transfers to the receiver on this step.}
\item
Finally, the state of the subsystem $B$ at the time instant $t$ is described by the marginal matrix
$\rho^B(\varphi,t)$,
\begin{eqnarray}\label{marg0}
\rho^B(\varphi,t)= {\mbox{Tr}}_{A,C} \rho(\varphi,t).
\end{eqnarray}
{{{}}  
As  pointed out in the Introduction, Sec.\ref{Sec:Introduction}, formula (\ref{marg0}) means 
that the final state of the whole system, in general, 
is not  a product state, i.e., $\rho(\varphi,t) \neq\rho^{AC}(\varphi,t)  \otimes  \rho^B(\varphi,t)$.  
Thus, in the real experiment, an
additional projection procedure is needed to  extract this state. 
}
\end{enumerate}
{{{}} 

We see that the first and the second steps of this algorithm are controllable. Both these steps 
serve to create the initial state of the whole system. The principal differences between them are  following.
\begin{enumerate}
\item
The first step is ''non-local'' because it involves all three subsystems. On the contrary,
the second step is local, it modifies the sender's initial state created on the first step. 
\item
The first step deals 
with the eigenvalues of all three subsystems, while the second step does not affect any eigenvalue. 
\item
The local parameters $\varphi_i$ may vary depending on the required receiver's state. In other words, they are 
the control parameters, unlike the eigenvalues  remaining unchanged during the state creation process.
\end{enumerate}
}

{{{}}  
From the above discussion it follows that the parameters of the 
local unitary transformation $U^A$  provide the tool allowing us to control 
the remote state creation 
process.  This control tool is {{}characterized} in next  Sec.\ref{Section:tool}. }

{{}  \subsection{
State creation with  
pure sender's initial state}
\label{Section:tool}}

{{{}}  First of all}, we shall note that not all $(N^A)^2-1$ parameters of the local unitary transformation 
$U^A\in~SU(N^A)$
 can affect  the state of receiver. 
We consider the map between the sender's control parameters  and the receiver's  creatable parameters 
in  the case of 
pure sender's initial state and 
deduce the number of effective control  parameters of $U^A$ (i.e., parameters which may really affect
 the receiver's state)
as a function of the sender dimensionality $N^A$.

It was shown  \cite{TBS,Z_QIP} that  $N^A-1$ 
parameters $\varphi_i$ (the number of independent eigenvalues) 
disappear from   initial state (\ref{instU}) {{}
because of the diagonality of the initial density matrix $\rho^A_0$.}     
In addition, the sender's initial density matrix  $\rho^A_0$ has a single non-zero eigenvalue (the pure state), 
so that
the number of variable  parameters of $U^A$ decreases by
$(N^A-1)(N^A-2)$
owing to the additional symmetry with respect to the transformation ${\mbox{diag}}\{1,\tilde U\}$,
$\tilde U\in SU(N^A-1)$. Thus, 
we   stay with
\begin{eqnarray}\label{DAA}
D^A=(N^A)^2-1 -(N^A-1) 
 -(N^A-1)(N^A-2) =2 (N^A-1)
\end{eqnarray}
parameters $\{\varphi_1,\dots,\varphi_{2(N^A-1)}\}$. 

Now we consider  evolution (\ref{ev})  governed by a certain Hamiltonian $H$. {{} Consequently,}
 the time $t$ appears as one more variable  parameter.
Thus, we have  $D^A+1 = 2 (N^A-1) +1$ 
variable real parameters of the sender
 which we refer to as the  control   parameters of the state creation algorithm. 
 
 {{{}}
In turn, the receiver's state of general position (\ref{marg0}) contains $(N^B)^2-1$ parameters which 
we refer to as the 
creatable parameters of the state creation algorithm. 
Therewith $N^B-1$ creatable parameters represent the independent eigenvalues 
$\lambda_i$ ($i=1,\dots,N^B-1$) of 
$\rho^B$,  while the rest $N^B(N^B-1)$ parameters $\beta_i$ 
appear in the eigenvector matrix
$U(\beta)$ of $\rho^B$, where all the parameters $\beta_i$ are collected in the list 
$\beta =\{\beta_1,\dots,\beta_{N^B(N^B-1)} \}$. 
Thus, the density matrix of the receiver's state
may be written as
 \begin{eqnarray}\label{rhoBf}
 \rho^B = U(\beta) \Lambda U^+(\beta) , \;\;\; \Lambda = {\mbox{diag}}(\lambda_1,\dots,\lambda_{N^B}), \;\;\;
 \sum_{i=1}^{N^B} \lambda_i =1.
 \end{eqnarray}
 }
As a result, we have the following 
map of $D^A+1$ control parameters of the sender's state  into $D^B$ creatable 
parameters of the  receiver's state:
\begin{eqnarray}\label{calM2}
{\cal{M}}(\varphi,t;\lambda,\beta): \; \{\varphi_1,\dots,\varphi_{2(N^A-1)}, t\} \to 
\{\lambda_1,\dots,\lambda_{N^B-1},\beta_1,\dots,\beta_{N^B(N^B-1)}\}.
\end{eqnarray}
We see that the number of variable parameters 
increases linearly with $N^A$ 
in the case of  pure sender's initial state. 

{{{}} In principle, 
since we consider a finite quantum system, all the creatable parameters $\lambda_i$ 
and $\beta_i$ may be analytically expressed in terms of the control parameters $\varphi_i$ and $t$. 
However, 
these expressions are very combersome even for small systems, so that the numerical consideration is 
a proper way of dealing with   map (\ref{calM2}).}

Obviously, we may hope to create the whole receiver's state-space  if  
\begin{eqnarray}\label{DADB}
{\mbox{the number of control parameters }}\;\; \ge \;\;{\mbox{the number of creatable  parameters }},
\end{eqnarray}
or, regarding map (\ref{calM2}),
\begin{eqnarray}\label{DADB1}
2 (N^A-1)+1\ge (N^B)^2-1 .
\end{eqnarray}
{{{}}
If inequality in (\ref{DADB}) is  strong (i.e., relation ''$>$'' is realized), then the time may be 
disregarded as a control parameter 
without destroying the validity of (\ref{DADB}).}
Consequently, we may expect to create the whole (or large) region of the receiver's state-space 
at a (properly) fixed
time instant $t_0$ {{{}}  (the time instant  $t_0$ is determined by the periodic behavior 
of the considered finite quantum system and will be found  in Sec.\ref{Section:4node} 
(see also Sec.\ref{Section:appendixC}) for a particular example)}.
Disregarding the time $t$ in  map (\ref{calM2}),
we reduce this map to the following one:
\begin{eqnarray}
\label{calM2red}
&&
{\cal{M}}(\varphi;\lambda,\beta): \; \{\varphi_1,\dots,\varphi_{2(N^A-1)}\} \to 
\{\lambda_1,\dots,\lambda_{N^B-1},\beta_1,\dots,\beta_{N^B(N^B-1)}\}.
\end{eqnarray}
{{}We shall note, that the only difference between maps (\ref{calM2})  and  (\ref{calM2red}) is the time $t$ in the list of 
control parameters of map (\ref{calM2}). However, because of this additional parameter, 
map (\ref{calM2}) may not be considered as a 
completely local one. In fact,} to obtain the required state of receiver, one has to transfer the 
information about the proper
time instant of the state registration  (classical channel). 
On the contrary, map (\ref{calM2red}) is
completely local because 
the receiver 
 registers  the state at  a fixed time instant $t_0$, which can be reported in advance. 
 {{{}}
 Note that the classical channel mentioned above 
 transmits the information about the registration time instant rather then the information 
 about the state itself as in the other state-creation algorithms 
 \cite{BDSSBW,BHLSW,PBGWK2,PBGWK,DLMRKBPVZBW,XLYG,G}.}

Below, in Sec.\ref{Section:examples}, we consider 
the  pure initial state of sender.    This choice is cased by the conclusion following from the 
numerical experiments with different initial states  $\rho^A(0)$. Namely, the maximal region of creatable 
states (or the creatable region) is associated with the pure sender's initial state.

{{{}}  Apparently,  the classical {{{}}  perfect  pure one-qubit} state transfer along
the spin chain \cite{Bose} may be considered as a very 
special case of the remote state creation process, 
see  Appendix \ref{Section:appendixA}.}

\section{Examples of  state creation in short homogeneous spin-1/2 chains}
\label{Section:examples}

{{{}}
\subsection{ Homogeneous  spin-1/2 chain governed by  nearest-neighbor XY Hamiltonian}
Let us consider the open spin-1/2 chain with the one-qubit receiver $B$ and the one- or two-qubit sender $A$. 
For definiteness, let $A$ and $B$ be placed, respectively, 
in the beginning and in the end of this chain. Therewith, the rest nodes of spin chain form the subsystem $C$ which we call the transmission line.  
Thus, we have a three-partite quantum system  $A-C-B$. For simplicity, 
only the one-qubit transmission line $C$ is considered here. Of course, such a
transmission line is a very 
short one, but, nevertheless, it enriches the features of the 
state-creation process.
We assume that the spin dynamics is governed by  
 the nearest-neighbor 
 $XY$ Hamiltonian $H$,
 \begin{eqnarray}\label{calH}
{{H}}= -\sum_{j=1}^{3} \frac{d}{2} (I^+_jI^-_{j+1} + I^-_jI^+_{j+1}). 
\end{eqnarray}
Here  $d$ is the coupling constant between the nearest neighbors, 
$I^\pm_j=I_{x;j} \pm i I_{y;j}$, $I_{\alpha;j}$, $\alpha=x,y,z$, are the projection operators  
of the  $i$th spin angular momentum. 
We put $d=1$ without the loss of generality.

Below we consider the state-creation  with the one- and two-qubit senders in more details. }

\subsection{Three-node chain with  one-qubit sender}
\label{Section:3node}

We proceed with the three node chain having the one-qubit subsystems $A$ (the 1st node), $B$ (the 3rd node) and  
$C$ (the 2nd node),
thus $N^A=N^B=2$. 
{{{}}   We  consider a pure initial states of the subsystems $A$ } and  $C$ and a mixed initial state of the 
receiver $B$.
Thus, the  initial state of the whole spin chain is given by expression (\ref{inst}) with
\begin{eqnarray}\label{rho0d}
\rho^A_0={\mbox{diag}}(1,0),\;\;\rho^C={\mbox{diag}}(1,0),\;\;\rho^B={\mbox{diag}}(\lambda^B,1-\lambda^B).
\end{eqnarray}
The unitary $SU(2)$  transformation $U^A$ responsible for the state creation 
is the two-parametric one:
\begin{eqnarray}\label{UA}
U^A(\varphi) &=&e^{-i \pi \varphi_2 \sigma_3}
e^{-i \frac{\pi \varphi_1}{2} \sigma_2}e^{i \pi \varphi_2 \sigma_3}=\left( \begin{array}{cc}
\cos\frac{\pi \varphi_1}{2} &-e^{-i 2 \pi\varphi_2}\sin\frac{\pi \varphi_1}{2} \cr
e^{i 2\pi \varphi_2}\sin\frac{\pi \varphi_1}{2} &\cos\frac{\pi \varphi_1}{2}
\end{array},
\right)
\\\label{int}
&& 0 < \varphi_i < 1,\;\;i=1,2,\;\;\varphi=\{\varphi_1,\varphi_2\}.
\end{eqnarray}
The evolution of this  chain is described by formula (\ref{ev}) with  Hamiltonian (\ref{calH})
  in accordance with the Liouville  equation. 
Finally, the state of the subsystem $B$ at some instant $t$  is described by the marginal matrix 
 $\rho^B(t)$ (\ref{marg0}),
 \begin{eqnarray}\label{marg1}
 \rho^B(t)=Tr_{A,C} \rho(t) = Tr_{A,C} e^{-i t H } \left(U^A \rho^A_0(U^A)^+\otimes  \rho^C_0
 \otimes  \rho^B_0\right) e^{i t H },
 \end{eqnarray}
which can be represented  in  form (\ref{rhoBf}) with
\begin{eqnarray}\label{UB}
\Lambda&=& {\mbox{diag}}(\lambda,1-\lambda),
\\\nonumber
U &=&e^{-i \pi \beta_2 \sigma_3}
e^{-i \frac{\pi \beta_1}{2} \sigma_2}e^{i \pi \beta_2 \sigma_3}
=\left( \begin{array}{cc}
\cos\frac{\pi \beta_1}{2} &-e^{-i 2\pi \beta_2}\sin\frac{\pi \beta_1}{2} \cr
e^{i 2\pi \beta_2}\sin\frac{\pi \beta_1}{2} &\cos\frac{\pi \beta_1}{2}
\end{array}
\right),
\\\label{intbet}\nonumber
&& 0 < \beta_i <1,\;\;i=1,2,\;\;\beta=\{\beta_1,\beta_2\},\;\;\frac{1}{2}\le \lambda \le 1.
\end{eqnarray}
Here the parameters $\beta_i$, $i=1,2$, and $\lambda$ depend on $\varphi$ and $t$, 
but we do not write them  as  arguments 
for the brevity.
Now  transformation (\ref{calM2})  reads
\begin{eqnarray}\label{calM2nodes3}
&&
{\cal{M}}(\varphi,t;\lambda,\beta): \; 
\{\varphi_1,\varphi_2, t\} \to 
\{\lambda,\beta_1,\beta_2\},\\
\label{inreg0}
&&
0\le \varphi_i \le 1,\;\;i=1,2,\;\; 0\le t \le \pi \sqrt{2},
\\\label{betlam0}
&&
0\le \beta_i \le 1,\;\;i=1,2, \;\;1/2\le \lambda \le 1.
\end{eqnarray}
{{{}}   It is remarkable that map (\ref{calM2nodes3})  admits a 
simplification due to the linear relation between
the parameters $\varphi_2$ and  $\beta_2$ (see Appendix \ref{Section:appendixB}), i.e., the  transformation 
$\varphi_2 \to \beta_2$
becomes trivial.}
This allows us to disregard the parameters $\varphi_2$ and $\beta_2$ in     
map (\ref{calM2nodes3}-\ref{betlam0}) and replace this map  with the following one:
\begin{eqnarray}\label{calM2nodes32}
&&
{\cal{M}}(\varphi_1,t;\lambda,\beta_1): \; 
\{\varphi_1, t\} \to 
\{\lambda,\beta_1\},\\
\label{inreg}
&&
0\le \varphi_1 \le 1,\;\; 0\le t \le \pi \sqrt{2},
\\\label{betlam}
&&
0\le \beta_1 \le 1, \;\;1/2\le \lambda \le 1, 
\end{eqnarray}
where we consider $\lambda\ge 1/2$ without the loss of generality. No new states may be created at $t>\pi\sqrt{2}$, 
which follows from the periodicity of the spin-dynamics 
{{{}}and is justified by the numerical simulations}.
Map (\ref{calM2nodes32}-\ref{betlam}) is numerically studied  in the following subsection.

\subsubsection{Numerical {{{}}   study of  state creation with one-qubit sender}}
\label{Section:3nodesnumerics}

We consider  map  (\ref{calM2nodes32}-\ref{betlam}) with initial condition (\ref{rho0d}).
The  creatable region in the 
space $(\lambda,\beta_1)$ (\ref{betlam})
is depicted in
Fig.\ref{Fig:beta12}a-d for the following set of $\lambda^B$:
\begin{eqnarray}\label{lamB}
\lambda^B =1,\;\;\frac{3}{4},\;\;\frac{1}{4},\;\;0. 
\end{eqnarray}
{{{}}  
Therewith, we use the following uniform splitting of the variation intervals  of the control parameters 
\begin{eqnarray}\label{uniform0}
&&
{\mbox{variation interval $[0,1]$ of $\varphi_1$  is splitted into 399 segments (400 points)}} \\\nonumber
&&
{\mbox{variation interval $[0,\pi\sqrt{2}]$ of $t$ is splitted into  2399 segments (2400  points)}}.
\end{eqnarray}
}
{{{}}Each point on this figure corresponds to a particular receiver's state. We see 
that these points form smooth lines and each of these lines  corresponds to a particular time instant of map 
(\ref{calM2nodes32})
with the parameter $\varphi_1$ running  
the interval $[0,1]$ specified in (\ref{inreg}).}
In the case of  pure initial state, $\lambda^B=1,0$, the lines cover the whole space $(\lambda,\beta_1)$ 
(see Fig.\ref{Fig:beta12}a,d). The vertical lines $\lambda=1$ in these figures are associated with
the time instant
$t=\pi\sqrt{2}$ corresponding to the perfect one-qubit pure state transfer from the first to the third node \cite{CDEL}. 
 The case {{{}}   of the  initial state with} $\lambda^B=1$ 
(Fig.\ref{Fig:beta12}a) is of the most interest because the map (\ref{calM2nodes32}) 
is mutually unique.  Moreover, 
the  lines are time-ordered in this case: the time instant prescribed to  each of these lines  increases in the 
direction of the arrows 
 from $0$ 
to $\pi\sqrt{2}$ {{{}}  (the dashed line with  arrows is not associated with the receiver's states)}. So, in principle, we may construct  the one-to-one relation between the pairs
($\varphi_1$, $t$) and ($\lambda$, $\beta_1$). Thus, having a particular pair 
($\lambda$, $\beta_1$), we may restore the 
parameter of sender $\varphi_1$ and the time instant  $t$ when the state was sent.
{{{}}Fig.\ref{Fig:beta12} is aimed to show the overall picture of the receiver's state distribution.  
Notice that the ''dense'' areas mean that more points from the sender's state-space 
are mapped into these  areas.

Regarding the mixed initial states $\lambda^B=3/4,1/4$, see Fig.\ref{Fig:beta12}b,c, 
not any state of the receiver
may be created by the local unitary transformations $U^A$, which is indicated  
by the  unavailable regions in these figures. 
In addition, the map (\ref{calM2nodes32}) is not mutually unique {{{}}(unlike the case $\lambda^B=1$, 
see Fig.\ref{Fig:beta12}a)} since
some particular  states ($\lambda$, $\beta_1$) can be created by more then one pair ($\varphi_1$, $t$)
{{{}} (because 
some lines cross each other)}.

We shall also note the case $\lambda^B=\frac{1}{2}$ when
{{{}}   only the states with $\frac{1}{2}\le \lambda\le 1$ and 
$\beta_1=0,1$ (i.e.,  arbitrary diagonal states) are creatable}.

\begin{figure*}
   \epsfig{file=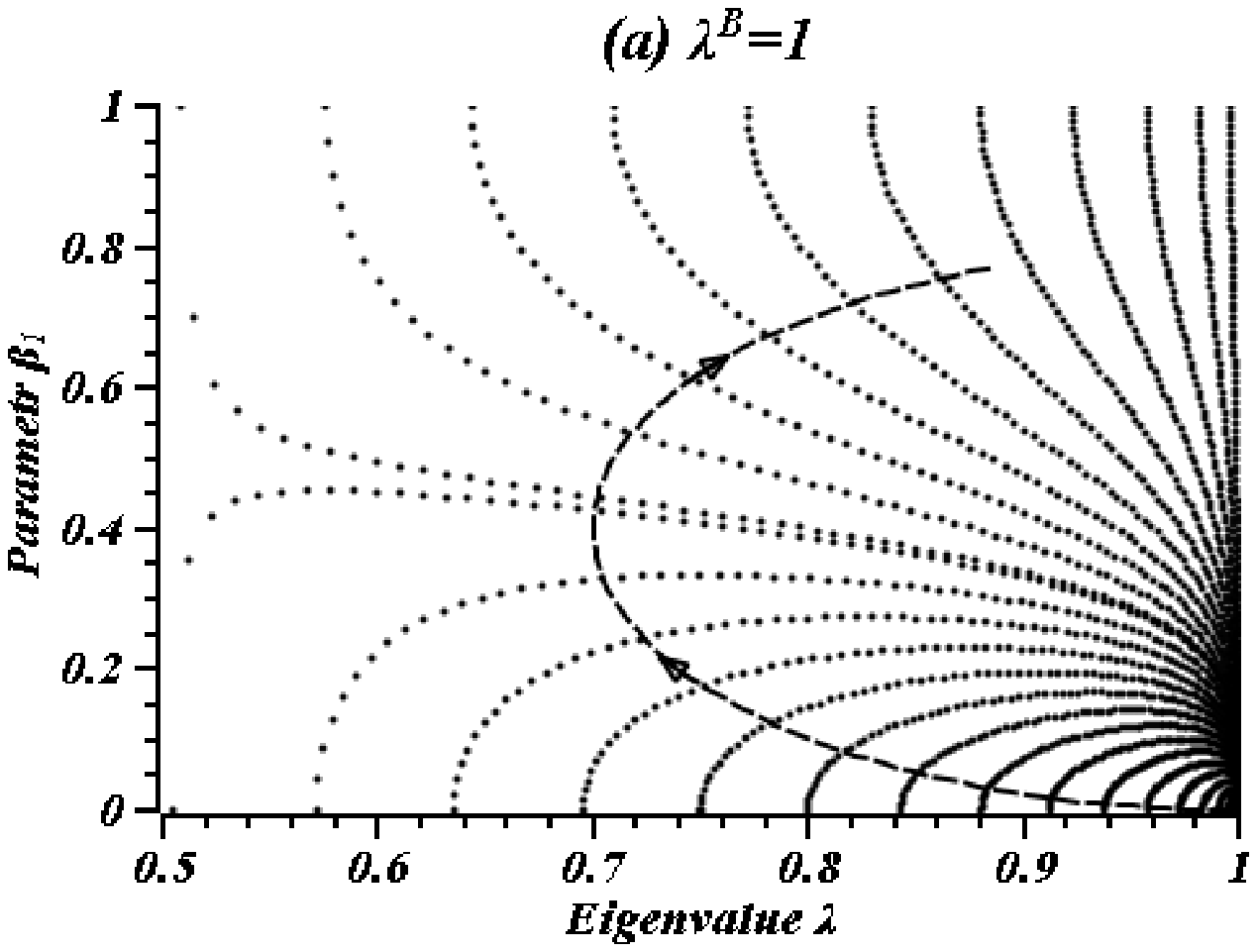,
  scale=0.6
   ,angle=0
}
\epsfig{file=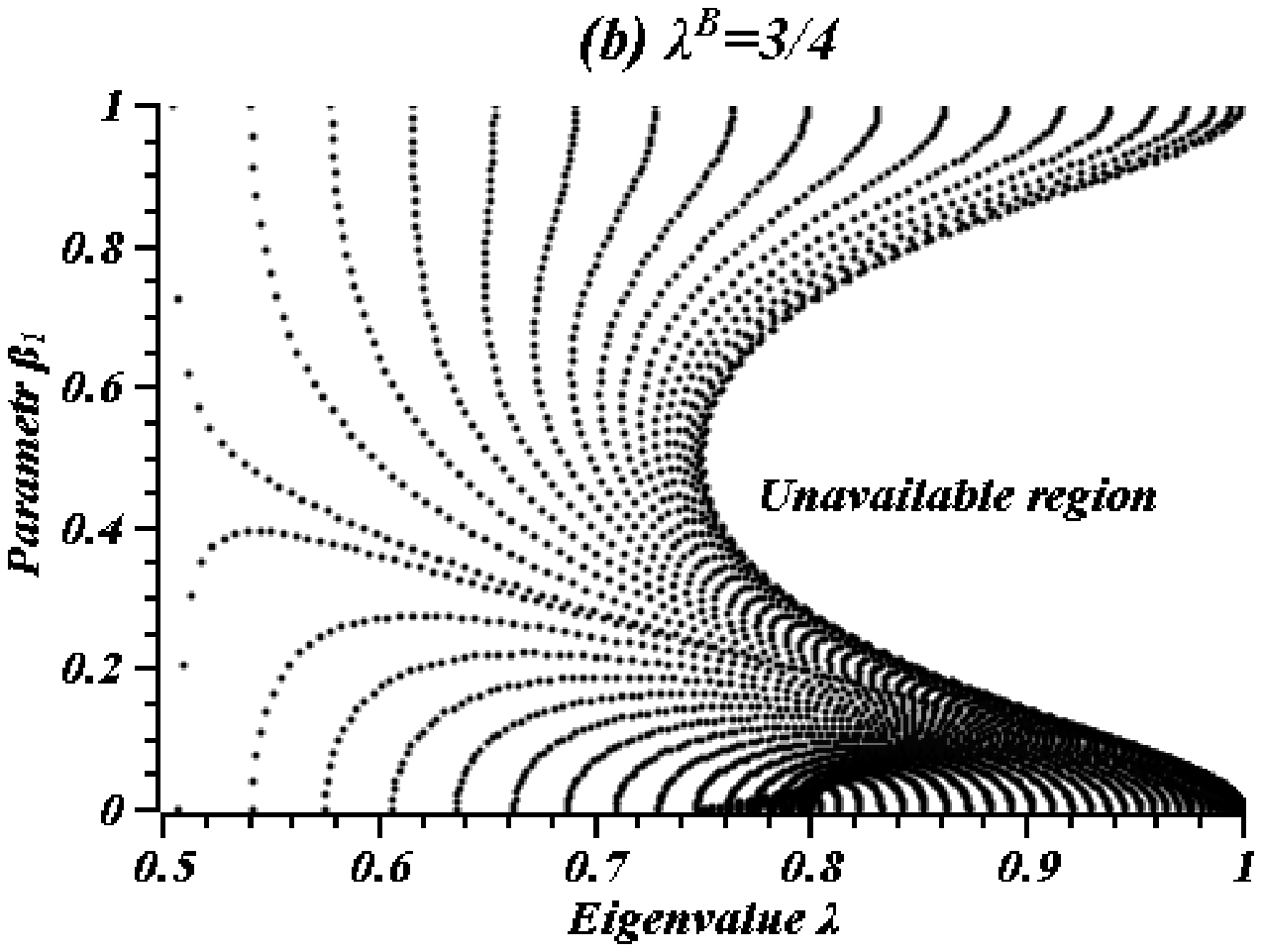,
  scale=0.6
   ,angle=0
}
 \epsfig{file=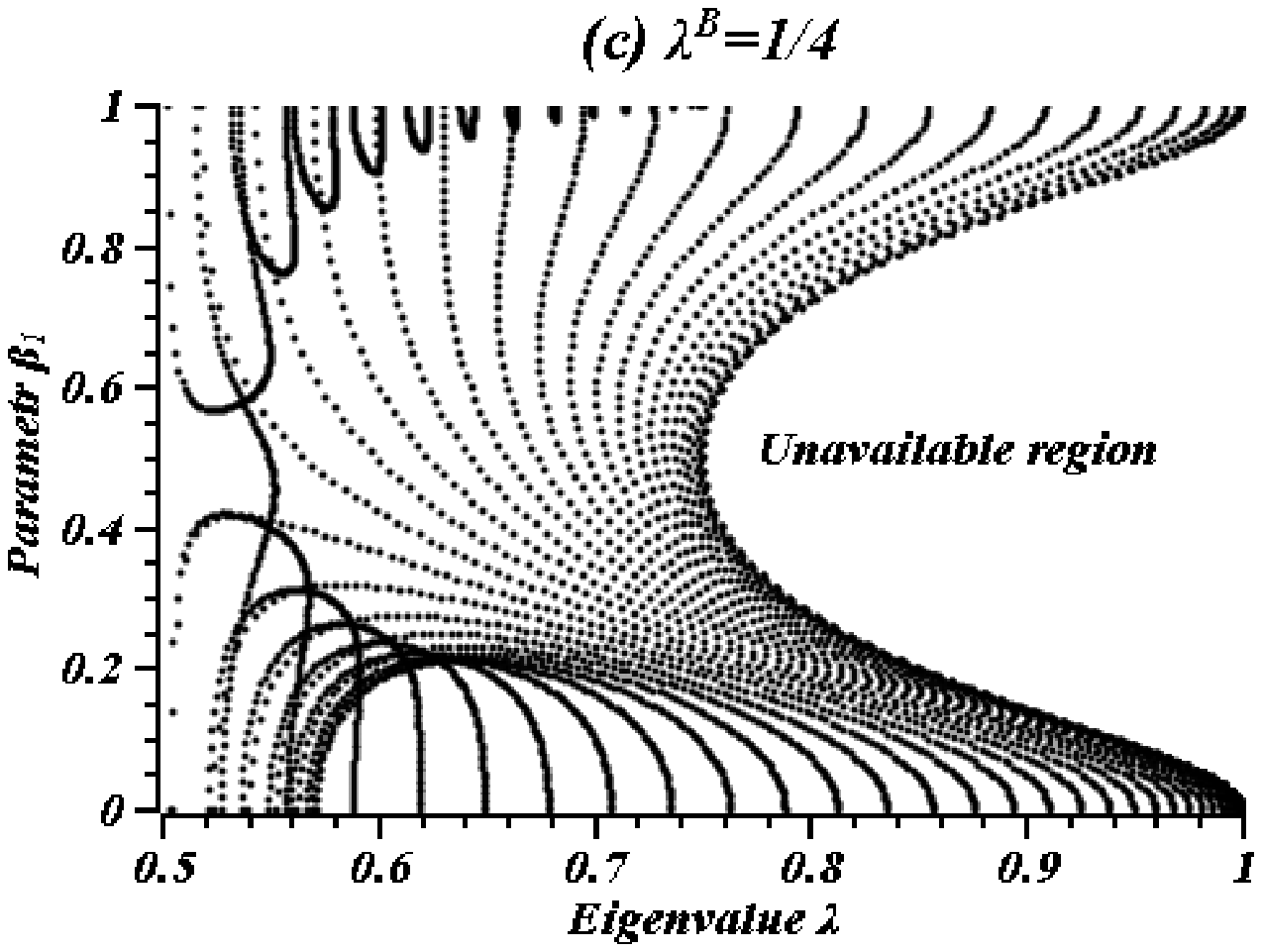,
  scale=0.6
   ,angle=0
}
\epsfig{file=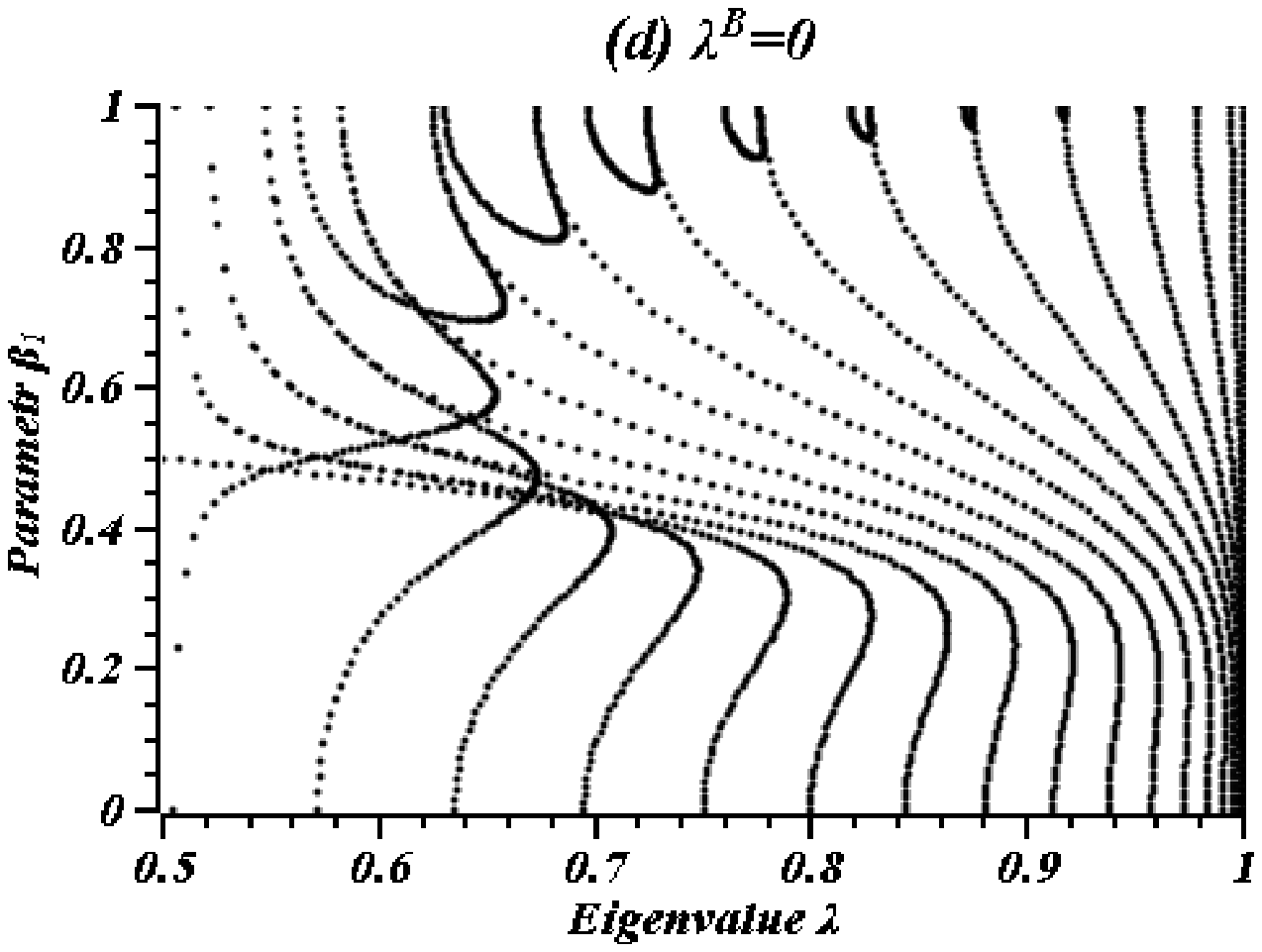,
  scale=0.6
   ,angle=0
}
\caption{The two-parameter receiver state-space $(\lambda,\beta_1)$ of  map (\ref{calM2nodes32}-\ref{betlam})
for the three node spin chain with the one-qubit sender
and set (\ref{lamB}) of $\lambda^B$ is considered.  
This figure demonstrates the non-uniform distribution of the creatable states.
(a) $\lambda^B=1$, the pure initial state; the whole space $(\lambda,\beta_1)$ is creatable and 
 map (\ref{calM2nodes32}) is mutually unique;
the time $t$ increases in the direction of the arrows. (b,c) $\lambda^B=3/4$ and  $\lambda^B=1/4$ respectively; 
the mixed initial state,  map (\ref{calM2nodes32}) is not mutually  unique and
the unavailable region appears. (d)  $\lambda^B=0$, 
the pure initial state; the whole space $(\lambda,\beta_1)$ is creatable, but  map  
(\ref{calM2nodes32}) is not mutually  unique.
} 
  \label{Fig:beta12} 
\end{figure*}

The overall disadvantage of the proposed algorithm of the state creation with 
equal dimensionalities of the sender and receiver is that 
 we have to 
involve the time   $t$  as a control parameter of map (\ref{calM2nodes32}) in order to cover a  valuable region of 
the space 
$(\lambda,\beta_1)$.
Consequently, this map is not completely governed by the local unitary transformation 
$U^A$.
 This disadvantage is compensated in the case $N^A=4$ considered in the next subsection.
Notice that the map (\ref{calM2nodes32}) with a pure initial state 
covers the complete state-space  $(\lambda,\beta_1)$ only in the case of 
two and three node chains with the nearest neighbor interactions. 
Involving  the dipole-dipole {{{}}  
interaction among all   nodes},  the unavailable region appears even 
in the case of {{{}}  three nodes 
and pure initial state. Besides, the unavailable region appears in the case of longer 
chains with nearest neighbor interactions
as well.  
In this regard, we have to remember the similar feature of the  
perfect one-qubit pure  state transfer}. Namely,  {{{}} the  
perfect one-qubit pure  state transfer  is possible  along the homogeneous spin-1/2  chains of
two and three nodes  governed by the 
nearest neighbor XY Hamiltonian \cite{CDEL},  and this phenomenon is destroyed
by involving all node interactions; the perfect state transfer is also    
impossible in longer  homogeneous chains with nearest neighbor interactions.
Thus, the remote  creation of the whole state-space  and perfect pure 
state transfer may be organized in the same chain, i.e., in the 
three-node homogeneous chain governed by  the nearest neighbor $XY$ Hamiltonian. However, it is not clear whether this is always valid.}
The  remote state creation
using the long non-homogeneous chain with the interaction constants providing the perfect one-qubit 
pure state transfer  
between the first and the last   nodes \cite{CDEL,KS} is not studied  here.

\subsubsection{Density function as a characteristics  of  creatable  {{} region}}
\label{Section:density}
The distribution of creatable states {{{}}   in the parameter space $(\lambda,\beta_1)$
 is non-uniform, see  Fig.\ref{Fig:beta12}a-b. This  is reflected in the varying 
density of points in this figure.} If we fix some small area in space 
$(\lambda,\beta_1)$, then the more points 
are in this area, 
the  more points from the space
$(\varphi_1,t$) are mapped into it.  
To better visualize  this effect, we introduce the so-called density function 
as follows:
\begin{eqnarray}\label{S}
&&
S(\lambda_i,\beta_{1j})=\lim_{N_\lambda,N_{\beta_1}\to\infty}
\frac{s(\lambda_i,\beta_{1j})}{ \varepsilon_\lambda\varepsilon_{\beta_1}N^{st}},\\\nonumber
&&
\lambda_i = \frac{1}{2}+\varepsilon_\lambda i,\;\;i=0,\dots, N_\lambda-1,\;\;
\beta_{1j} = \varepsilon_{\beta_1} j,\;\;j=0,\dots, N_{\beta_1}-1,\;\;
\varepsilon_\lambda =\frac{1}{2  N_\lambda},\;\;\varepsilon_{\beta_1} =\frac{1}{  N_{\beta_1}},
\end{eqnarray}
where $s(\lambda_i,\beta_{1j})$ is the number of states in the rectangle
\begin{eqnarray}\label{regepsilon}
 (\lambda_i,  \lambda_i+\varepsilon_\lambda],\;\;
 (\beta_j,\beta_j+\varepsilon_\beta],
 \end{eqnarray}
 $N^{st}$ is the total number of states in the rectangle (\ref{betlam}).
 The function $S$ is normalized as follows:
 \begin{eqnarray}\label{norm}
 \sum_{i=0}^{N_\lambda-1}\sum_{j=0}^{N_{\beta_1}-1} S (\lambda_i,\beta_{1j})\varepsilon_\lambda\varepsilon_{\beta_1} = 1.
 \end{eqnarray}
We represent the contour plot of the density function (\ref{S}) in Fig.\ref{Fig:3nodesdensity} for
$\varepsilon_{\beta_1}=2\varepsilon_\lambda=\varepsilon=0.01$ and $N^{st}=960000$.
 This value of $N^{st}$ {{{}}   corresponds to}
the uniform splitting  (\ref{uniform0}) of the variation  intervals (\ref{inreg}) {{{}}   of}
the parameters 
$\varphi_1$ and $t$. 
Figs.\ref{Fig:3nodesdensity}a-d  show   the  dependence of the density function 
on $\lambda^B$. The choice of $\lambda^B$ must be  defined  by a particular 
area of states which we need to create. {{}Perhaps, the bright areas   are of most interest.}
The maximal values $S_{max}$ of the density function 
 together with their  coordinates $\lambda_{max}$ and $\beta_{1;max}$  
 for different values of $\lambda^B$ from  list (\ref{lamB}) are collected in Table \ref{Table1}.

\begin{figure*}
   \epsfig{file=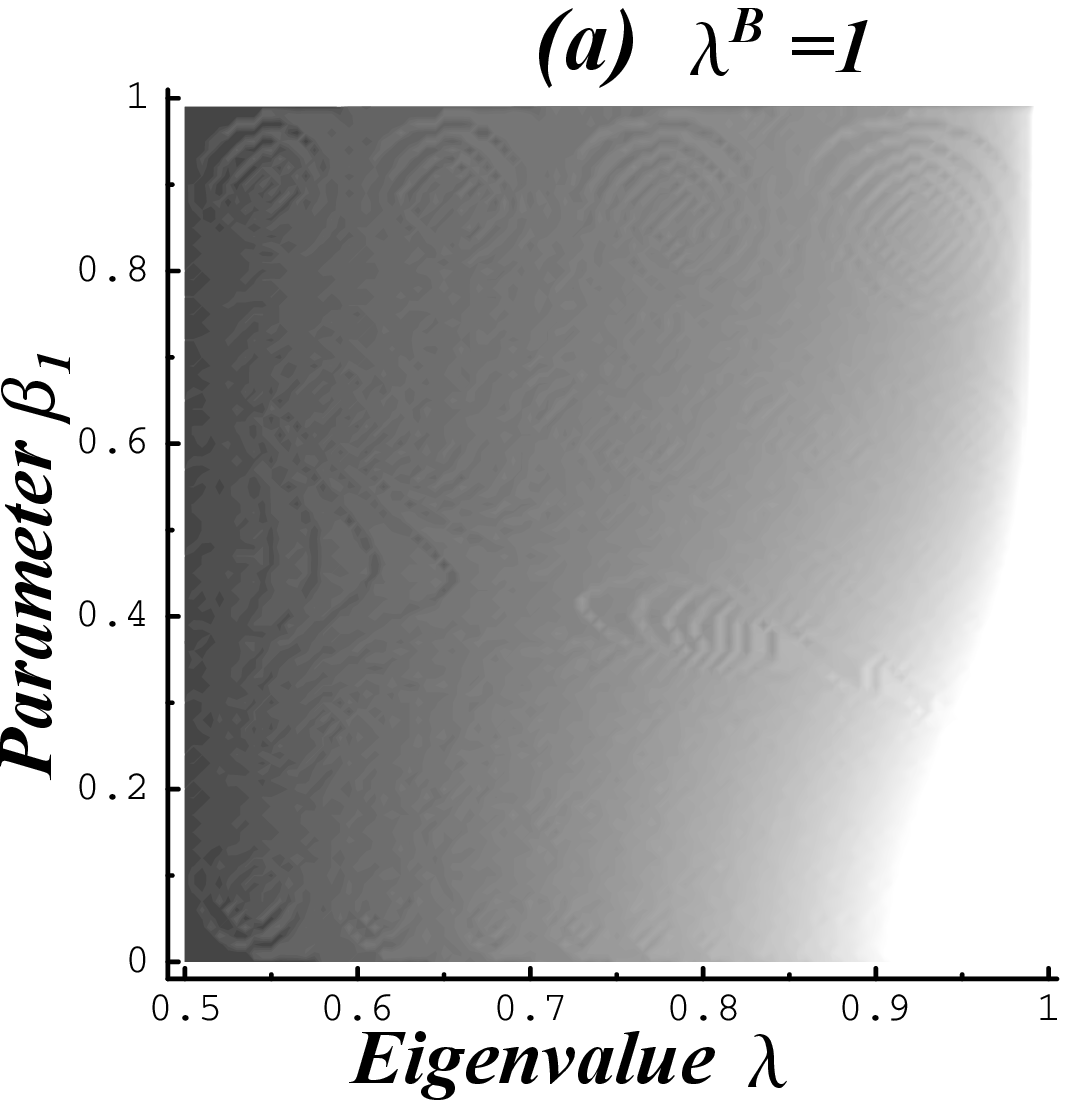,
  scale=0.6
   ,angle=0
}
 \epsfig{file=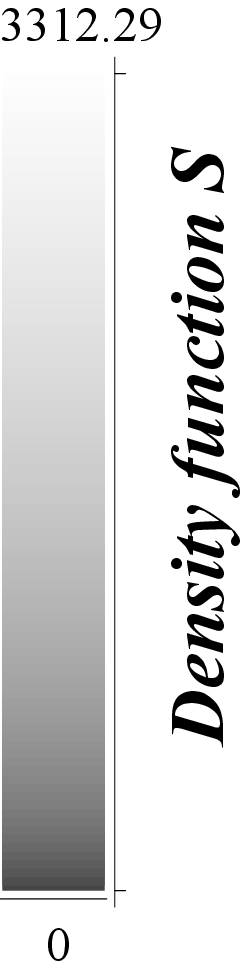, scale=0.6
   ,angle=0
}
   \epsfig{file=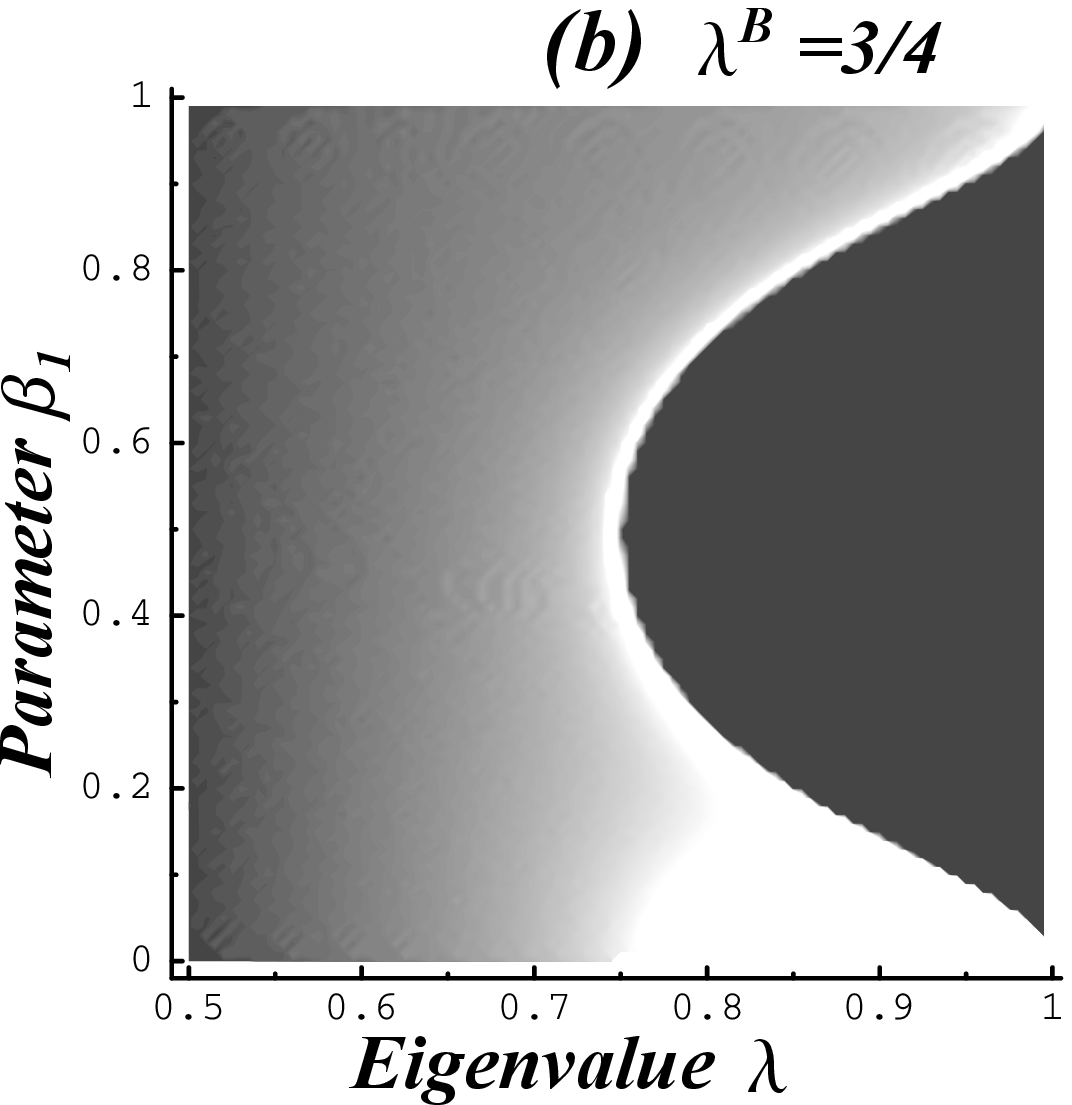,
  scale=0.6
   ,angle=0
}
 \epsfig{file=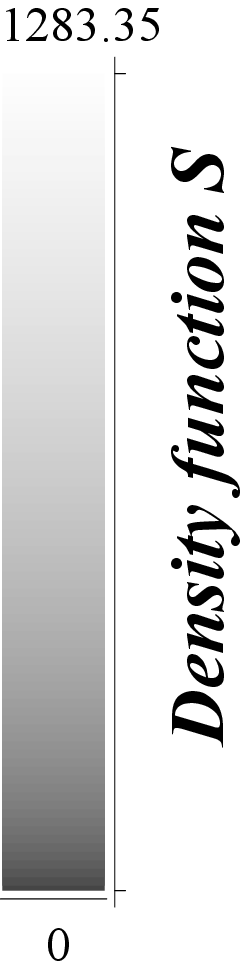, scale=0.63312
   ,angle=0
}
\newline
   \epsfig{file=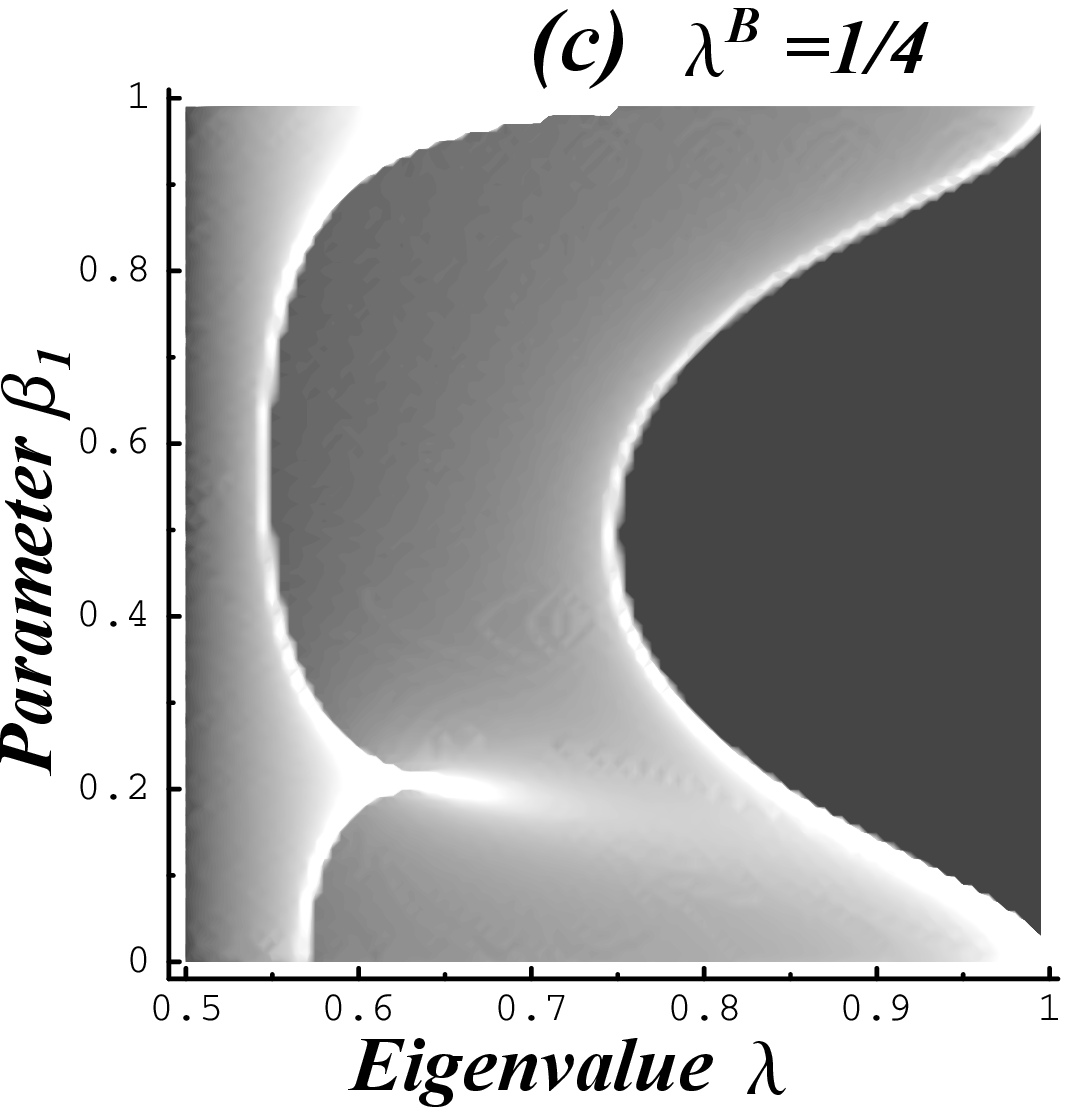,
  scale=0.6
   ,angle=0
}
 \epsfig{file=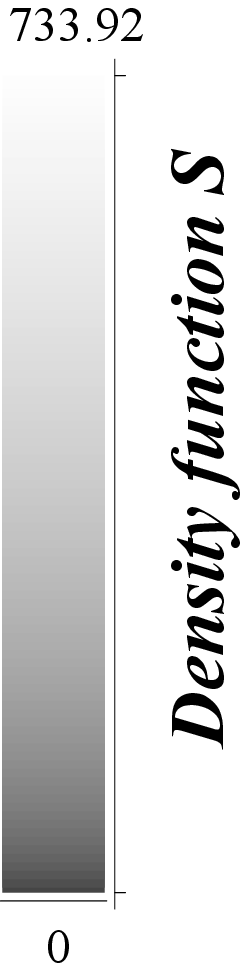, scale=0.6
   ,angle=0
}
   \epsfig{file=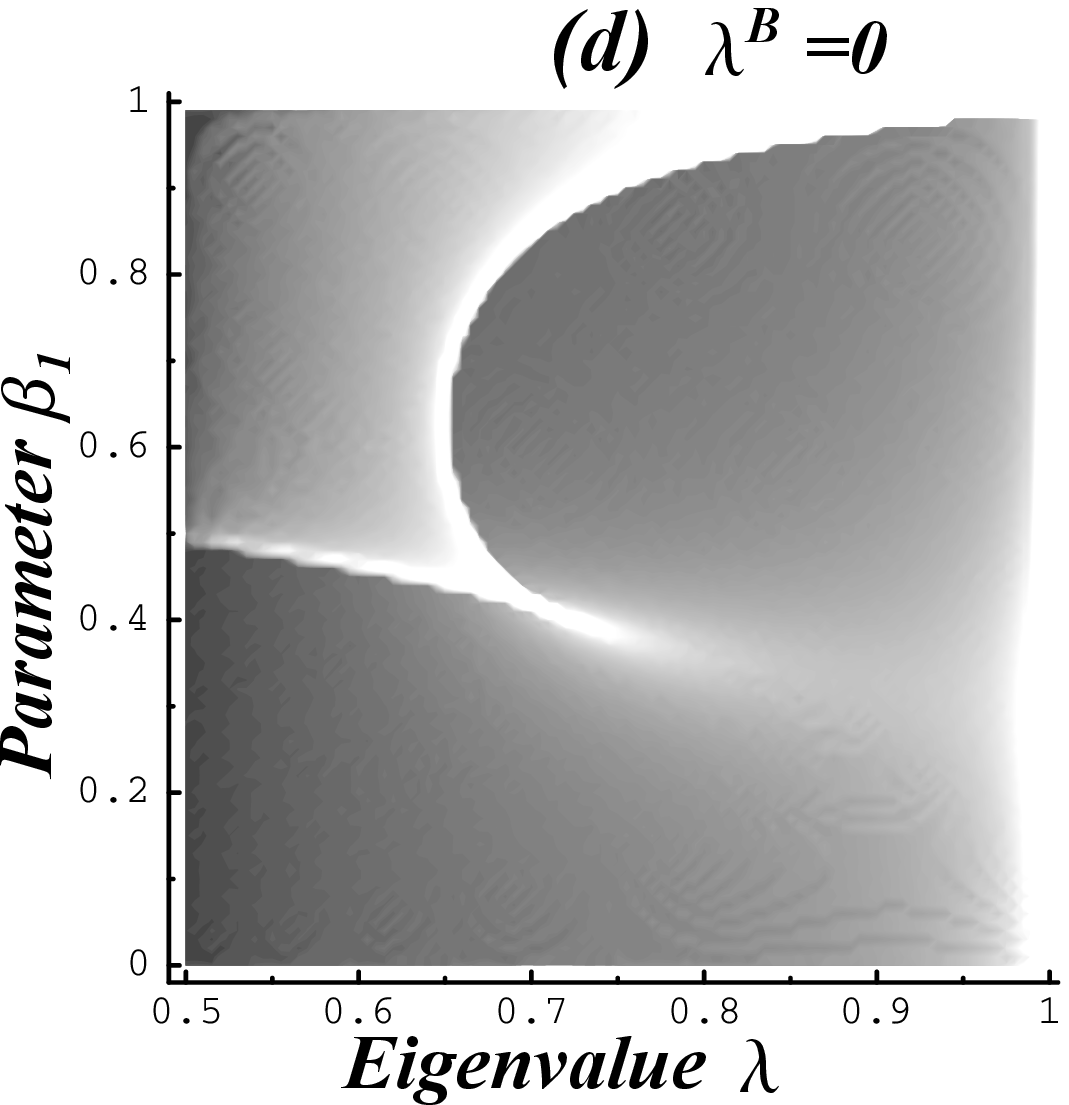,
  scale=0.6
   ,angle=0
}
 \epsfig{file=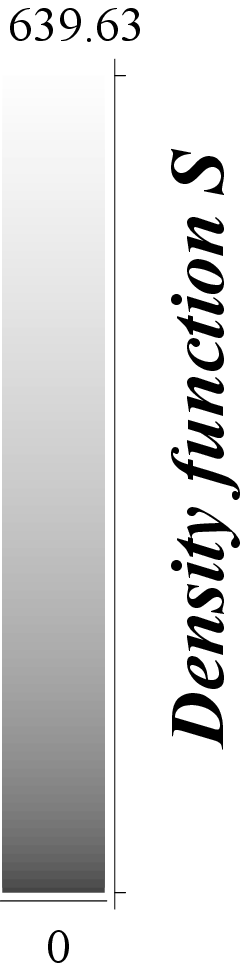,
  scale=0.6
   ,angle=0
}
\caption{The density function of the two-parameter receiver state-space $(\lambda,\beta_1)$ 
of  map (\ref{calM2nodes32}-\ref{betlam}) for the three node  chain with 
the one-qubit sender and set (\ref{lamB}) of $\lambda^B$ is considered ( 
compare with Fig.\ref{Fig:beta12}).  
{{}The bright areas are most ''populated''}. 
The maximal values of the density function $S_{max}$ together with the appropriate values of the parameters
$\lambda_{max}$ and $\beta_{1;max}$ are collected in Table \ref{Table1}.
(a, d) $\lambda^B=1$ and $0$ respectively, the pure initial state; 
the whole receiver's state-space   may be created. (b, c)  $\lambda^B=3/4$ and $1/4$ respectively, 
the mixed initial state;  the whole receiver's state-space  may not be created;  
the unavailable regions are indicated by
the black areas in  both figures.} 
  \label{Fig:3nodesdensity} 
\end{figure*}

\begin{table}
\begin{tabular}{|c|cccc|}
\hline
 $\lambda^B$ & $1$ & $\displaystyle\frac{3}{4} $& $\displaystyle\frac{1}{4}$ & $0$\cr \hline
 $S_{max}$ &3312.29&1283.35&733.92&639.63\cr
 $\lambda_{max}$ &0.9975& 0.7525&0.7475&0.9975\cr
 $\beta_{1;max}$ & 0.005&0.005&0.995&0.995\cr\hline
\end{tabular}
\caption{The maximal values $S_{max}(\lambda_{max},\beta_{1;max})$ 
of the  density function of creatable states   for the case of one-qubit sender (the three node chain). 
{{{}}   {\it Remark:}
One can show that the values $S_{max}$ increase with decrease of $\varepsilon$ in accordance with 
definition of the density function (\ref{S}), 
whereas the coordinates 
$\lambda_{max}$ and $\beta_{1;max}$ remain unchanged  up to the accuracy, respectively, 
$\pm  \varepsilon/2$ and 
$\pm  \varepsilon$.
Thus, the values in the 1st line are underestimated. The same remark holds for Table \ref{Table2}.
}}
\label{Table1}
\end{table}

Although the density function $S$ illustrates the distribution of the creatable states, this distribution is 
also understandable from  Fig.\ref{Fig:beta12} because the creatable states are arranged in lines.
The case of two-qubit sender {{{}}(studied in Sec. \ref{Section:4node})} is different. 
The creatable states are not arranged in the well-structured lines, so that the 
density function $S$ becomes very important in that case.

{{{}}  
\subsection{Four-node spin chain with  two-qubit sender}
\label{Section:4node}
}

Let us consider the four node chain with the  two-qubit sender (the first and the second nodes), 
while the receiver (the 4th node) and the transmission line (the 3rd node)
remain one-qubit subsystems. {{{}}
{{} As was shown in \cite{TBS,Z_QIP} (see also the beginning of Sec.\ref{Section:tool}),
 the local unitary transformation 
$U^A\in SU(4)$ has  12 control parameters in the case of diagonal sender's initial state $\rho^A_0$:}}
\begin{eqnarray}\label{UrhoU}
&&U^A(\varphi)=\\\nonumber
&&
e^{i\pi \gamma_3 \varphi_1}e^{i\gamma_2 \frac{\pi\varphi_2}{2}}
e^{i \pi \gamma_3 \varphi_3}e^{i\gamma_5 \frac{\pi \varphi_4}{2}}
e^{i \pi\gamma_3 \varphi_5}e^{i\gamma_{10} \frac{\pi \varphi_6}{2}}
e^{i\pi \gamma_3 \varphi_7}e^{i\gamma_2 \frac{\pi \varphi_8}{2}}e^{i \pi \gamma_3 \varphi_9}
e^{i\gamma_5 \frac{\pi \varphi_{10}}{2}}e^{i\pi \gamma_{3} \varphi_{11}}e^{i\gamma_2 \frac{\pi \varphi_{12}}{2}},
\\\label{varphi2}\label{phi_reg}
&&
\varphi=\{\varphi_1,\dots,\varphi_{12}\},
\;\;
0\le \varphi_i \le 1,\;\;i=1,\dots,12.
\end{eqnarray}
The explicit matrix representation  of $\gamma_i$ is given in Appendix \ref{Section:appendixC} \cite{TBS,GM}. 

We consider the diagonal initial density matrix of form (\ref{inst}) and 
fix the pure initial state of the subsystems $A$  and $C$, while the initial
state of the receiver $B$ is arbitrary diagonal one, similar to Sec.\ref{Section:3node}.
In other words, our initial matrices are  following:
\begin{eqnarray}\label{inmatrices}
\rho_0^A={\mbox{diag}}(1,0,0,0),\;\;
\rho_0^C={\mbox{diag}}(1,0),\;\;\rho_0^B={\mbox{diag}}(\lambda^B,1-\lambda^B).
\end{eqnarray}
The density matrix evolves in accordance with  eq.(\ref{ev}).
Finally, the marginal matrix $\rho^B(t)$ describing the receiver's state may be calculated using formulas 
(\ref{marg1},\ref{UB}) of Sec.\ref{Section:3node}, where  the Hamiltonian $H$ is the same
(see eq. (\ref{calH})), while  
 $U^A$ and $\rho^{A,B,C}_0$ are given by eqs.(\ref{UrhoU}) and (\ref{inmatrices}), respectively.

In accordance with Sec.\ref{Section:stcr},
in this case,  we may disregard the time as a varying parameter of the state creation 
and use   map (\ref{calM2red}) which now has 6 effective arbitrary parameters in accordance with 
eq.(\ref{DAA}).
However, since we need to create the three parametric  receiver's state, 
it might be enough to take just three parameters $\varphi_i$ in the map (\ref{calM2red}). 
{{{}}  
But these three parameters may not be chosen in an arbitrary way and, in general, 
the choice of parameters affects  the
creatable region.}
The preferable choice of three parameters is not evident, and the 
problem of the optimal {{{}}  
(i.e., leading to the maximal possible creatable region)}
parametrization of  the considered map remains beyond the scope of this paper. 
Instead, we propose the following parametrization which yields  large 
({{{}}and, perhaps, the maximal  possible creatable region for the given initial state 
(\ref{inmatrices})}). {{{}}   At least, the numerical experiments with a set of 
other   parameterizations give the same result, see also Appendix \ref{Section:appendixD}.}

{{{}}In our parametrization,}
the parameters $\varphi_{10}$, $\varphi_{11}$ and $\varphi_{12}$ vary independently, 
while all other parameters are 
linearly expressed in terms of the single parameter $\phi$: 
\begin{eqnarray}
&&
\varphi_{2n-1} = \phi,\;\;n=1,\dots,5,\\\nonumber
&&
\varphi_{2n} =\frac{\phi}{2},\;\;n=1,\dots,4.
\end{eqnarray}
Thus, the map (\ref{calM2red})
simplifies to
\begin{eqnarray}\label{2M4nodes}
{\cal{M}}(\varphi,\beta)&:&  \{\varphi_{10},\varphi_{11},
\varphi_{12},\phi\}\to \{\lambda,\beta_1,\beta_2\},\\\label{2varphi4nodes}
&&
0\le \varphi_i\le 1,\;\;i=10,11,12,\;\;0\le \phi \le 1,\\\label{2beta4nodes}
&&
0\le \beta_i\le 1,\;\;i=1,2,\;\;\frac{1}{2} \le \lambda  \le 1,
\end{eqnarray}
which is numerically studied in the following subsection.
Therewith, we  fix a time instant $t_0$ inside of some interval,
\begin{eqnarray}
\label{timeint}
0< t_0 \le T,
\end{eqnarray}
{{{}}
where the parameter $T$ is 
related with the periodicity of  quantum dynamics of the 
considered finite spin system. It is closely related with the minimal (in magnitude) 
eigenvalue $\lambda^{abs}_{min}$: $T\approx \frac{\pi}{\lambda^{abs}_{min}}$ 
(half of the oscillation period of the real and imaginary parts of the exponent 
$e^{i\lambda^{abs}_{min}t}$). }
{{{}}
We have $\lambda^{abs}_{min}\approx 0.309$, so that  $T\approx 10.2$.  One can show that 
during this time interval 
the evolution of the area of  creatable region } 
{{{}}  passes through its maximal value}

\subsubsection{Numerical {{{}}   study of state creation with two-qubit sender}}
\label{Section:numerics}
{{{}}   The numerical simulation show that the distribution of the 
receiver's states is uniform in $\beta_2$, similarly to Sec.\ref{Section:3node} 
(see Appendix \ref{Section:appendixD} for details). 
Perhaps,} this means that there is a linear relation between $\beta_2$ and a particular parameter of $U^A$, 
similar to the relation between the parameters $\beta_2$ and $\varphi_2$ in Sec.\ref{Section:3node}.
However, we do not establish such a relation.
We only remove $\beta_2$ from the right side of map (\ref{2M4nodes}) and study the two-parameter state-space 
$(\lambda,\beta_1)$ of the receiver: 
{{{}}  
\begin{eqnarray}\label{2M4nodessimpl}
{\cal{M}}(\varphi,\beta)&:&  \{\varphi_{10},\varphi_{11},
\varphi_{12},\phi\}\to \{\lambda,\beta_1\} 
.
\end{eqnarray}
}
As explained in the Appendix, see the end of Sec.\ref{Section:appendixD}, 
 we  choose the time instant $t_0=6.4$ inside of 
 interval (\ref{timeint}) and perform 
{{{}}
a set of state-creating experiments substituting different values 
of $\lambda^B$ from  set (\ref{lamB}) into
 initial condition 
(\ref{inmatrices}).}
{{{}}  Therewith we use the following  uniform 
splitting of the variation intervals $[0,1]$ of the parameters:
\begin{eqnarray}\label{splitting}
&&
{\mbox{variation intervals of }} \;\;\varphi_{10,12}  \;\;\;{\mbox{are splitted  into 50 segments (51 points)}},\\\nonumber
&&
{\mbox{variation intervals of }} \;\;\varphi_{11}, \phi  \;\;\;{\mbox{are splitted into 25 segments (26 points)}}.
\end{eqnarray}
}
In all these cases, there is a 
boundary in the space $(\lambda,\beta_1)$ separating the creatable and unavailable regions, see also 
Fig.\ref{Fig:fournodes}a,c in  Appendix \ref{Section:appendixD}. {{{}} Perhaps, both 
the presence 
of the unavailable regions for all $\lambda^B$ and  the fact that the perfect pure state transfer 
is impossible along
the homogeneous four node spin chain} {{{}}   governed by the 
nearest-neighbor XY Hamiltonian \cite{CDEL} have the same origin.}  

{{{}}  To simplify 
the representation of  obtained  results, we show only the boundaries 
corresponding to the different values of $\lambda^B$ instead of the creatable regions themselves 
and depicture all these boundaries in the same figure, see 
Fig.\ref{Fig:fournode_creatable_reg}a}. Herewith the creatable region is to the left from the appropriate boundary 
line,
while the unavailable region is to the right from it.

We see
that the largest creatable region corresponds to the pure initial  states ($\lambda^B=1,0$).
It is important that there  is a small region 
which may not be created in the chain with  initial state (\ref{inmatrices}) 
and any choice of $\lambda^B$ at the selected time instant $t_0$. 
This absolutely unavailable region is indicated in Fig.\ref{Fig:fournode_creatable_reg}a and it is 
shown
in 
Fig.\ref{Fig:fournode_creatable_reg}b using the
proper scale. 
However, perhaps this region is creatable using  another initial state {{{}} 
and/or different time instant $t_0$}. 
It is interesting to note that the  numerically obtained boundaries  in Fig.\ref{Fig:fournode_creatable_reg}b can
be approximated by {{{}} the following analytical curves}:
\begin{eqnarray}\label{ancurve}
 0.9914 - 2.0034 (0.9999 - x)^{0.28}
&& {\mbox{the upper curve}},\\\nonumber
-0.0100 + 2.0369 (1 - x)^{0.28}&& {\mbox{the lower curve}}.
\end{eqnarray}
Similar to  Sec.\ref{Section:3nodesnumerics}, the case $\lambda^B=\frac{1}{2}$ 
allows us to create only the arbitrary diagonal states of the receiver, i.e., $\frac{1}{2}\le \lambda \le 1$ and 
$\beta_1=0,1$.

\begin{figure*}
   \epsfig{file=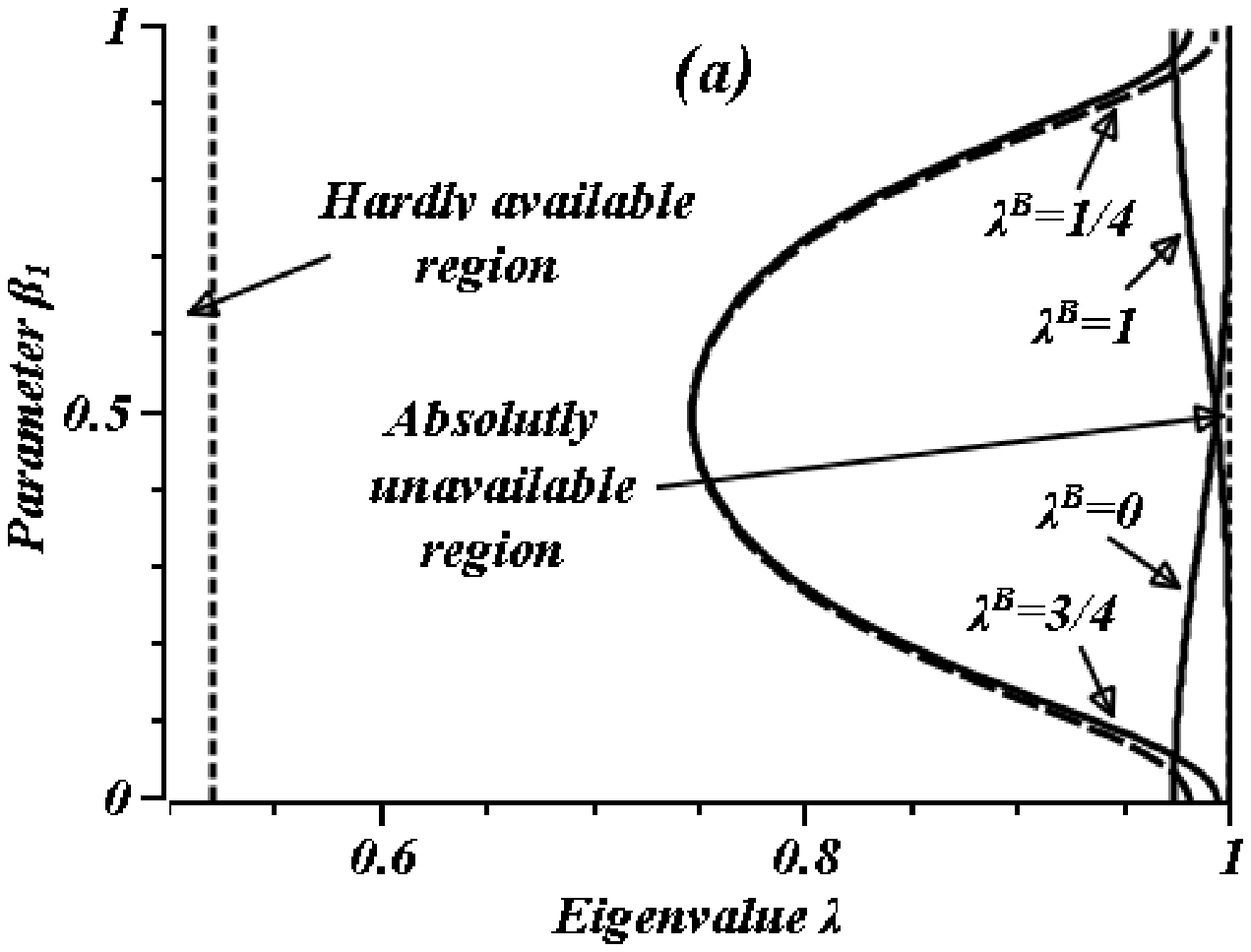,
  scale=0.6
   ,angle=0
   },
   \epsfig{file=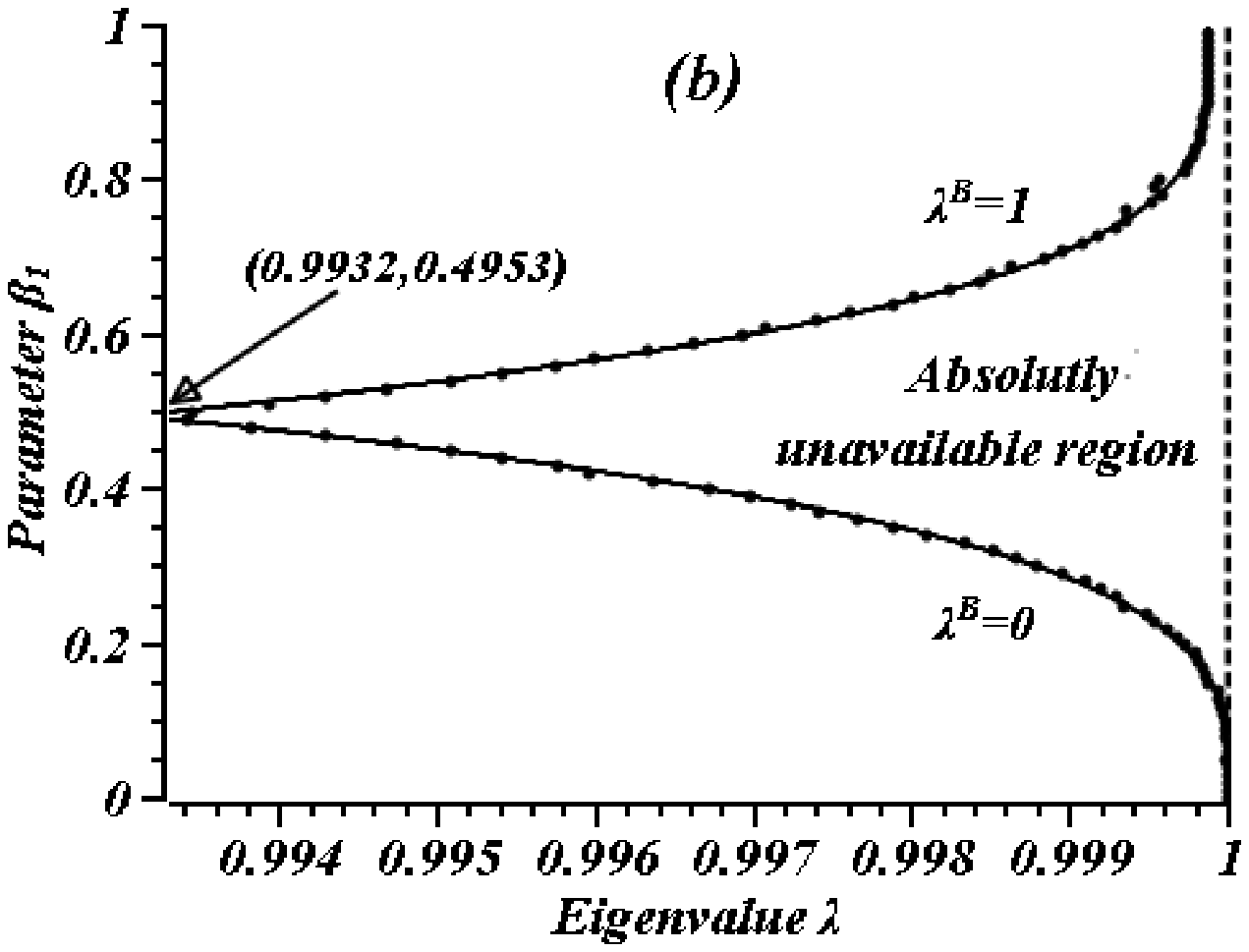,
  scale=0.6
   ,angle=0
}
\caption{The two-parameter space $(\lambda,\beta_1)$ 
of  map (\ref{2M4nodes}-\ref{2beta4nodes}) for the four node  chain with the
two-qubit sender and  set (\ref{lamB}) of  $\lambda^B$ is considered at $t_0=6.4$.  
We represent the boundary curves separating the creatable and unavailable regions of 
the receiver's state-space.
The creatable region is to the left from the appropriate boundary curve. 
There is a  region which may not be created by any 
local unitary transformation of
the subsystem $A$ {{{}}with any value of $\lambda^B$} (the absolutely unavailable region). 
The vertical stripe of states  with $\frac{1}{2}\le\lambda\le 0.52$ 
is hardly creatable, see also 
Figs.\ref{Fig:density}  and \ref{Fig:fournodes}.
($a$) The creatable and unavailable  regions of the receiver's state-space  for $\lambda^B=1,3/4,1/4,0$. ($b$) 
The absolutely unavailable region is bounded by the boundaries corresponding to the pure initial states 
$\lambda^B=1$, $0$; the
solid lines represent   analytical curves
(\ref{ancurve}).
} 
  \label{Fig:fournode_creatable_reg} 
\end{figure*}

\subsubsection{Density function of  creatable {{}region}}

To characterize the effectiveness of the state creation,   we 
use the   density function $S(\lambda,\beta_1)$ introduced in Sec.\ref{Section:density}.
Herewith, for each  $\lambda^B$ from  set  (\ref{lamB}), we construct the family of all  creatable  
states varying the parameters of the local transformation $U^A$ inside of the region (\ref{2varphi4nodes}).
We represent the contour plot of the density function  in Fig.\ref{Fig:density} for
$\varepsilon_{\beta_1}=2\varepsilon_\lambda=\varepsilon=0.01$ and $N^{st}=1758276$. 
This value of $N^{st}$ 
 is related with   uniform  splitting {{{}}  (\ref{splitting})
 of  variation intervals (\ref{2varphi4nodes}) 
of the parameters 
$\varphi_{10,12}$  and $\varphi_{11}$, $\phi$.}
The creatable region is maximal for the pure initial states ($\lambda^B=1,0$), as is shown in
Figs.\ref{Fig:3nodesdensity}a,d. 
The maximal values $S_{max}$ of the density function 
 together with their  coordinates $\lambda_{max}$ and $\beta_{1;max}$ for different  $\lambda^B$ from  set 
 (\ref{lamB}) are collected in Table \ref{Table2}.

\begin{figure*}
  \hspace{-0.5cm}  \epsfig{file=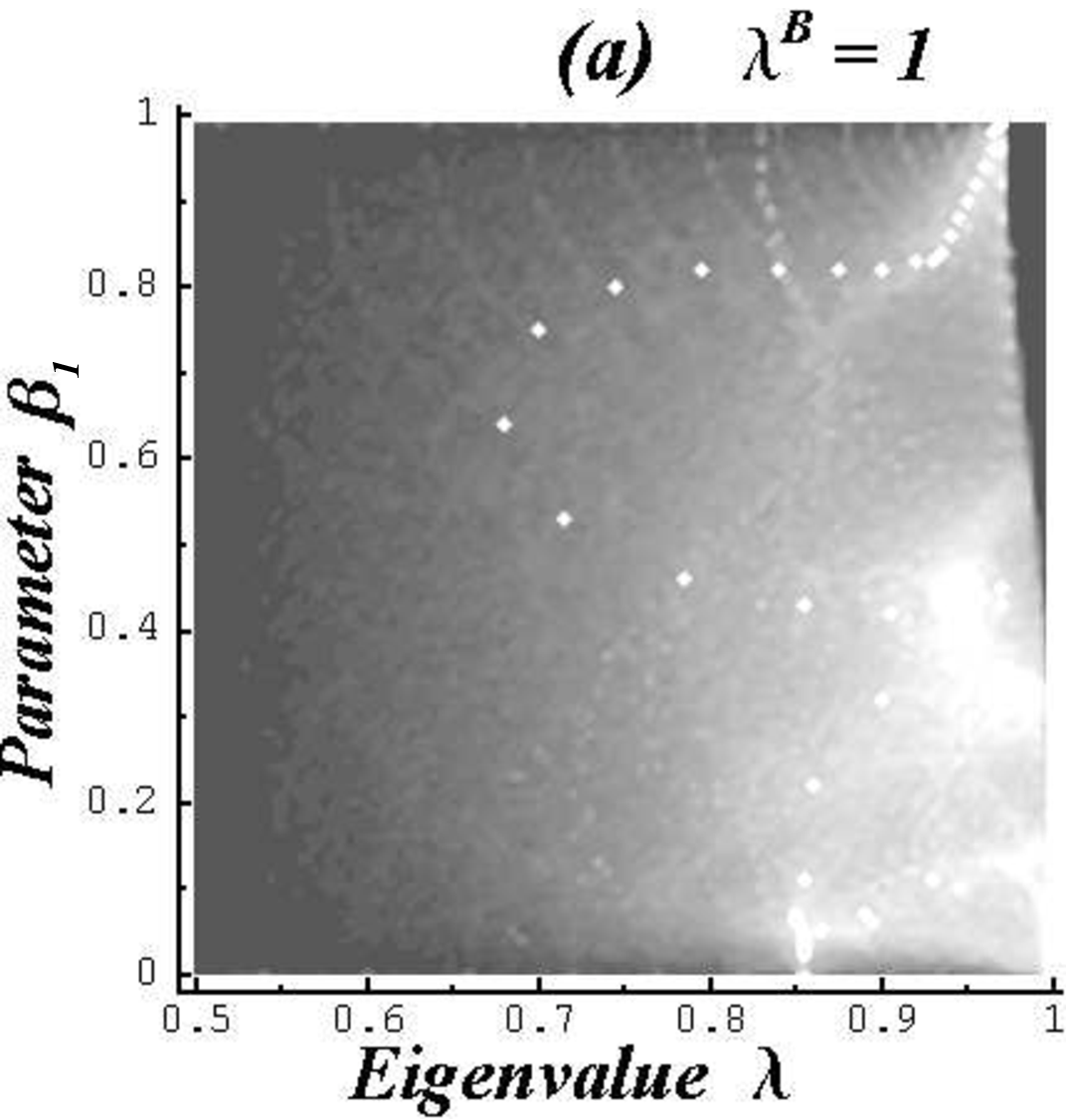,
  scale=0.6
   ,angle=0
}
 \epsfig{file=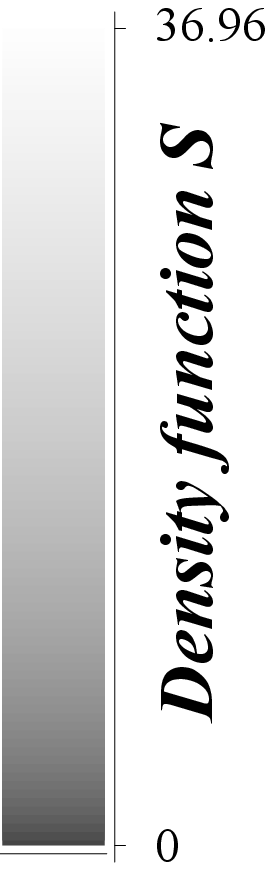, scale=0.6
   ,angle=0
}
   \epsfig{file=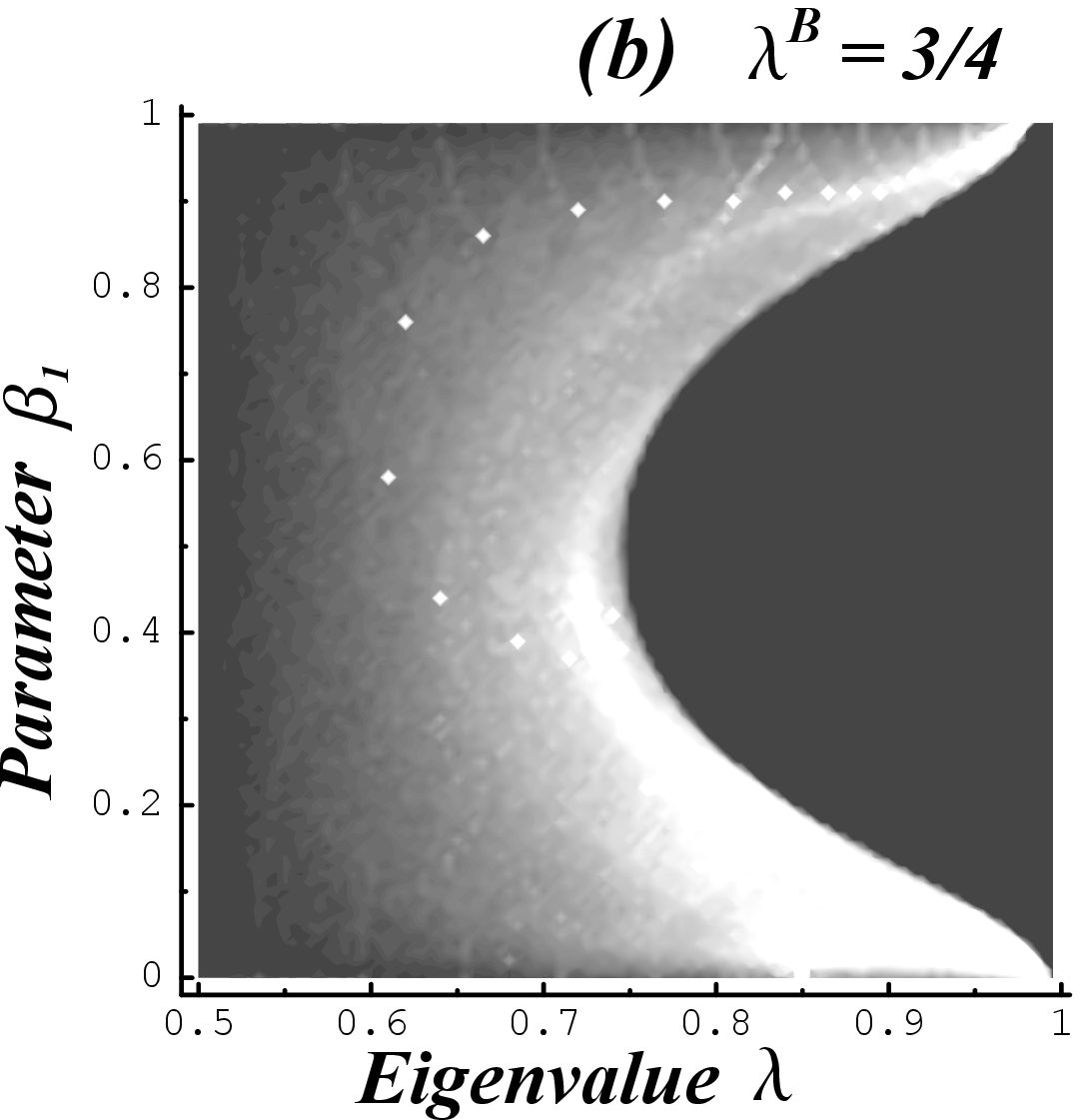,
  scale=0.6
   ,angle=0
}
 \epsfig{file=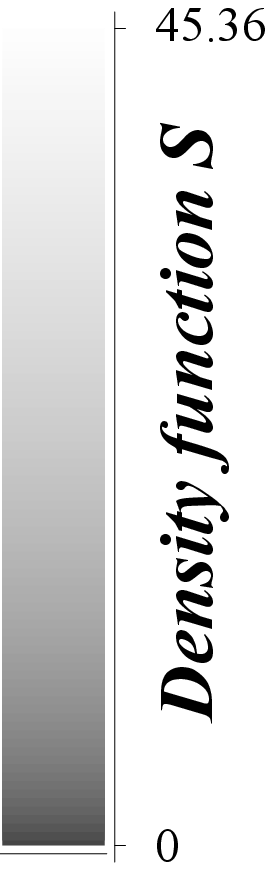, scale=0.6
   ,angle=0
}
\newline
 \hspace*{-0.5cm} 
 \epsfig{file=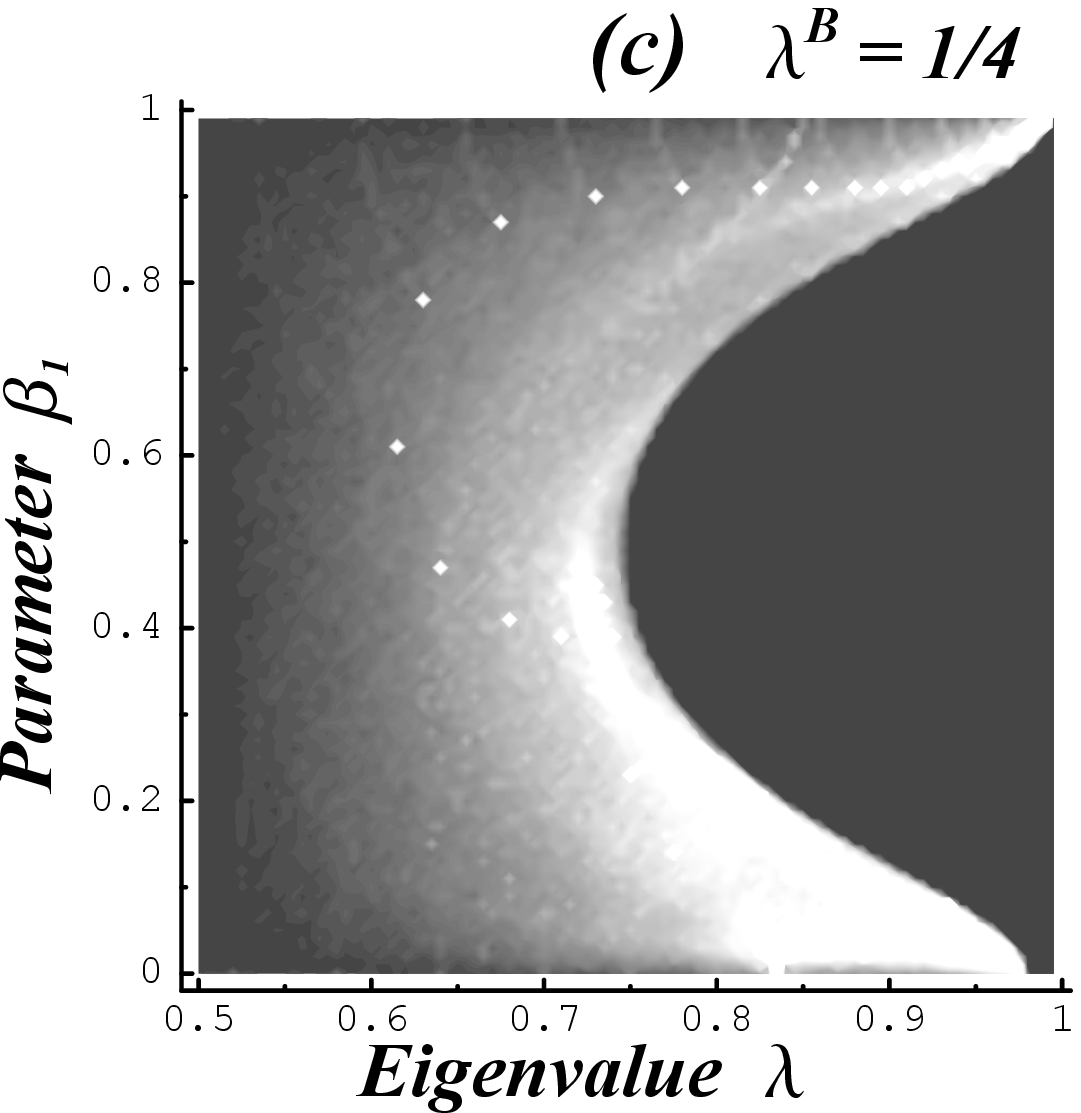,
  scale=0.6
   ,angle=0
}
 \epsfig{file=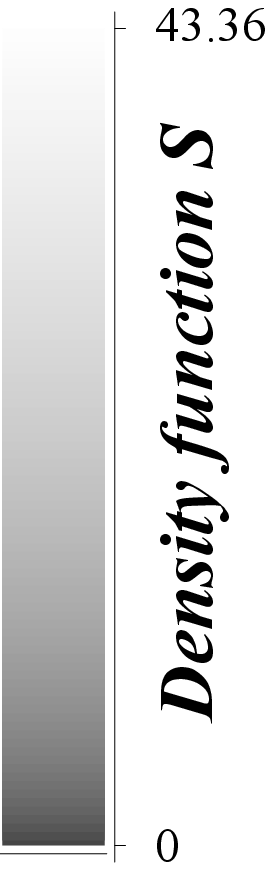, scale=0.6
   ,angle=0
}
   \epsfig{file=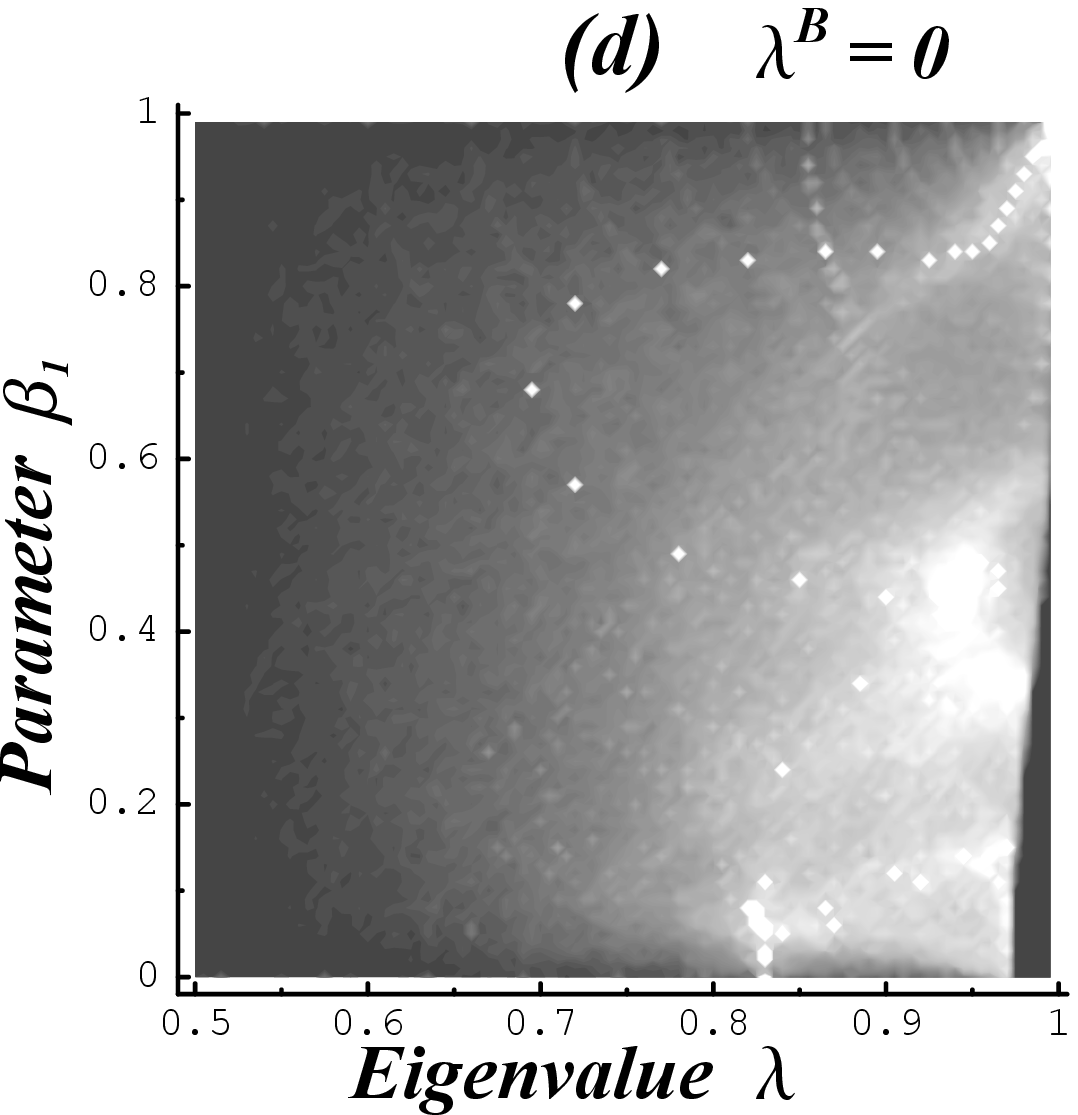,
  scale=0.6
   ,angle=0
}
 \epsfig{file=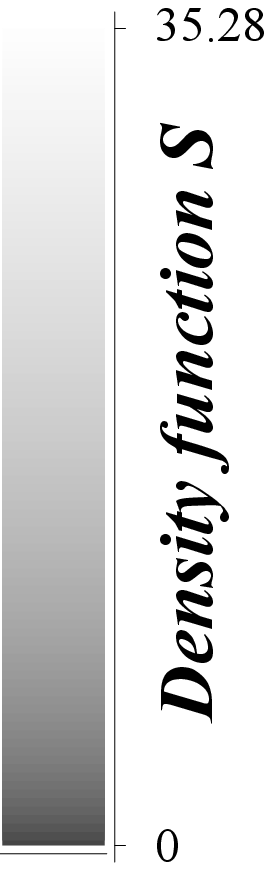,
  scale=0.6
   ,angle=0
}
\caption{ The density function of the two-parameter receiver state-space $(\lambda,\beta_1)$ 
of  map (\ref{2M4nodes}-\ref{2beta4nodes}) for the four node  chain with the
two-qubit sender and  set (\ref{lamB}) of $\lambda^B$ is considered.  
{{}The bright areas to the left 
from the right boundaries of the creatable regions are most ''populated''.} 
The maximal values of the density function $S_{max}$ together with the appropriate values of the parameters
$\lambda_{max}$ and $\beta_{1;max}$ are collected in Table \ref{Table2}.
The family of bright spots 
embedded into the dark areas appears for all $\lambda^B$. 
(a, d) $\lambda^B=1$ and $0$ respectively, the pure initial state; 
unlike the chain with the one-qubit sender,
the unavailable region (the black area)  appears even  in this case. (b, c)  $
\lambda^B=3/4$ and $1/4$ respectively, 
the mixed initial state; 
the unavailable regions are indicated by
the black areas, these regions are similar to those shown 
in Figs. \ref{Fig:3nodesdensity}b,c.} 
  \label{Fig:density} 
\end{figure*}

\begin{table}
\begin{tabular}{|c|cccc|}
\hline
 $\lambda^B$ & $1$ & $\displaystyle\frac{3}{4} $&  $\displaystyle\frac{1}{4}$ & $0$\cr\hline
 $S_{max}$ &36.96&45.36&43.36&35.28\cr
 $\lambda_{max}$ &0.8525&0.8475&0.8275&0.9975\cr
 $\beta_{1;max}$ &0.065&0.025&0.035&0.995\cr\hline
\end{tabular}
\caption{The maximal values $S_{max}(\lambda_{max},\beta_{1;max})$ 
of the  density function of creatable states for the case of  two-qubit sender (the four node chain)}
\label{Table2}
\end{table}

The unavailable regions (the black right sides) and the hardly creatable regions 
(the dark left sides) are well depicted in these figures.
{{} The most ''populated'' areas correspond to the bright parts of the figures. }
We shall point out the set of  bright spots 
inside of the dark area in each of  figures (a)-(d).
Perhaps, with increase of $N^{st}$, these spots join with each 
other forming a  smooth 
line.  The position of these sports
depends on a particular 
choice of the parameters $\varphi_i$ in  map (\ref{calM2red}) 
realizing the state creation (like the parameters $\varphi_i$, $i=10,11,12$, and $\phi$ in 
map (\ref{2M4nodes}-\ref{2beta4nodes})).
The meaning {{{}} and application} of these  spots are not clear. 

{{{}}  
\section{Conclusions  and discussion} 
\label{Section:Conclusion}}
We study the problem of remote creation of the mixed receiver's  states using 
the variable local unitary 
transformations of the sender and  assuming that the receiver's state  may not 
be locally transformed. Therewith,  the 
dynamics of the whole quantum system  is governed  by a particular  Hamiltonian. 
Since this problem is multi-parameter and very complicated, we proceed with the detailed study 
of  the state creation of  the  one-qubit receiver in the short spin-1/2 chains governed by the nearest-neighbor  
$XY$ Hamiltonian using the local unitary transformations of the either 
one-(the three node chain, $N^A=N^B=2$) or two-qubit (the four node chain, $N^A=4$, $N^B=2$) sender. 
We show that the time $t$ 
is an important control parameter in the case $N^A=N^B=2$ 
allowing us to cover the whole state-space of the receiver.  On the contrary, having $N^A=4>N^B=2$,
we may effectively control the state creation using  
only the local unitary transformations
of the sender provided that the time instant for the state registration at the receiver side is 
properly fixed. {{{}}   Here ''properly'' means that the creatable region of the sender's state-space 
must be large enough to involve the state we are interested in.}
We think that the  case $N^A>N^B$  is most promising 
because all possibilities of the receiver's 
state control are collected in the sender's side, so that the remote 
state-creation becomes really locally controlled without any classical 
communication between the sender and receiver.

The density function of the creatable states reflects the effectiveness of our algorithm. This function 
is used
in the both
cases of one- and
two-qubit senders. In the two-qubit sender case, Fig.\ref{Fig:density}, the set of bright 
(high density) spots  embedded into the dark region appears in the graphs. The positions of these spots 
depend on the particular choice of the 
parameters $\varphi_i$ used for the state-creation. 
The meaning and  possible application of such spots are not known yet. 

{{{}}  
An interesting results is the presence of the unavailable region of the receiver's states, i.e.,
such region
of states which  
may not be created by the initial local unitary transformations of the sender. 
These regions depend on  the initial state of the whole system. 
A possible benefit of unavailable regions is the sharing of the receiver's state-space among several senders 
having different (non-overlapping) creatable regions in the whole space of the receiver's states. 
This effect deserves a special study.
}

{{{}}  Notice also that using the three 
node homogeneous spin-1/2 chain  (Sec.\ref{Section:3node}) with the pure 
initial state we can create the whole state-space of the one-qubit  receiver 
(i.e., there is no unavailable region in this case, see Figs.\ref{Fig:beta12}a,d). Meanwhile,  
 {{{}}  the same chain allows the perfect  one-qubit pure state transfer 
 \cite{CDEL}}. Thus, these two phenomenon, perhaps, have the same origin.  But it 
is not clear whether
we may avoid the appearance of the 
unavailable  region in the  state-space of the receiver considering the
longer chain with the parameters providing  the perfect one-qubit pure state transfer \cite{CDEL,KS}.
}

{{{}}   
Finally, we would like to mention  the set of  numerical experiments 
demonstrating that the unavailable 
region of the receiver's state-space 
increases if we either  increase  the length of the chain or 
 replace the nearest node interaction 
with the 
all  node interaction. We do not represent results of these calculations. }

Author thanks Prof. E.B.Fel'dman and Dr. S.I.Doronin for helpful comments. 
{{} Author is also grateful to the referee for the detailed and useful criticisms of this manuscript.}
This work is partially supported by the Program of the Presidium of RAS No.8 
''Development of methods of obtaining chemical compounds and creation of new materials'' 
and by the RFBR grant No.13-03-00017

\section{Appendix}
\label{Section:appendix}
 {{{}}  
 \subsection{{{} Perfect} pure state transfer as a special case of state creation}
 \label{Section:appendixA}
 The arbitrary one-qubit pure state transfer along the spin chain \cite{Bose} 
 may be considered as a very special case of 
 the state creation via map (\ref{calM2}), when all the parameters of the local 
 unitary transformation $U^A$ are fixed. Having in mind the spin-1/2 
system in a strong homogeneous magnetic field, 
we replace  Hamiltonian (\ref{calH}) with the following one
\begin{eqnarray}\label{H}
 &&{\cal{H}}={{H}} + \gamma I_z,\;\;I_z=\sum_{i=1}^N I_{z;i},
\end{eqnarray}
where $H$ is given in eq.(\ref{calH}), $I_{z;i}$ is the $z$-projection operator
of the $i$th-spin angular momentum, and $\gamma$ 
is the Larmor frequency of the global external magnetic field.  The last term  in eq.(\ref{H}), describing the 
 interaction with the homogeneous  external magnetic field, was not important in Secs.\ref{Section:examples} 
 because its effect can be embedded into the 
 local unitary transformation $U^A$. But now, since $U^A$ is fixed, $\gamma$ becomes an important parameter
 and the map 
 (\ref{calM2}) must be replaced with the following one:
 \begin{eqnarray}\label{calMsttr}
&&
{\cal{M}}(\gamma,t;\lambda,\beta): \; \{\gamma,t\} \to 
\{\lambda_1\equiv\lambda, \beta_1,\beta_2\},\\\nonumber
&&
0\le \beta_i\le 1,\;\;i=1,2, \;\;\lambda=1.
\end{eqnarray}
Herewith, by the definition of the perfect state transfer, the fixed 
parameters $\varphi_1$ and $\varphi_2$ of the given  pure initial state of 
sender must be equal, respectively, to 
the parameters $\beta_1$ and $\beta_2$  of the receiver:
\begin{eqnarray}
\varphi_i=\beta_i,\;\;i=1,2.
\end{eqnarray}
 We see that there are only two control parameters in the map (\ref{calMsttr}) which must create 
three required values of the parameters  $\lambda$, $\beta_1$ and $\beta_2$.
 This  is impossible in the  long homogeneous chain. 
For this reason, to realize the pure state transfer, we need 
additional efforts, namely, the rigorous adjustment of the parameters of  the spin chain, 
such as the  interaction constants \cite{CDEL,KS}
and/or the local Larmor frequencies \cite{DZ}. Such an  adjustment, generally speaking, 
is not required in the mixed state-creation process considered in this paper, when the local unitary transformations serve 
to create the needed receiver's state.
}

\subsection{ Reduction of  map (\ref{calM2nodes3}-\ref{betlam0}) to (\ref{calM2nodes32}-\ref{betlam}) }
\label{Section:appendixB}
Since   initial state (\ref{rho0d})  is diagonal,
 using $U^A$ given by expression (\ref{UB}), we may write
(remember, that $I_{z;i} = \frac{1}{2} \sigma_3$)
\begin{eqnarray}
&&
\rho_0(\varphi)=U^A\rho_0^A (U^A)^+\otimes \rho_0^C\otimes \rho_0^B=\\\nonumber
&&
e^{-i \pi \varphi_2 \sigma_3}
e^{-i \frac{\pi \varphi_1}{2} \sigma_2}\rho_0e^{i \frac{\pi \varphi_1}{2} \sigma_2}
e^{i \pi \varphi_2 \sigma_3}
 \otimes
e^{-i \pi \varphi_2 \sigma_3} \rho_0^C e^{i \pi \varphi_2 \sigma_3}\otimes
e^{-i \pi \varphi_2 \sigma_3} \rho_0^B e^{i \pi \varphi_2 \sigma_3}=\\\nonumber
&&
e^{-i 2 \pi\varphi_2 I_z}
\left(e^{-i \frac{\pi \varphi_1}{2} \sigma_2}\rho_0e^{i \frac{\pi \varphi_1}{2} \sigma_2}
\otimes \rho_0^C \otimes \rho_0^B  \right)  
e^{i 2 \pi\varphi_2 I_z},
\end{eqnarray}
where
\begin{eqnarray}
\label{III}
e^{i 2 \pi\varphi_2 I_z}=e^{i 2 \pi\varphi_2 I_{z;1}} \otimes e^{i 2 \pi\varphi_2 I_{z;2}} 
\otimes  e^{i 2 \pi\varphi_2 I_{z;3}}.
\end{eqnarray}
Since $[H,I_z]=0$, the evolution of the density matrix reads 
\begin{eqnarray}\label{ev2}
\rho(t)= e^{-i H t} \rho_0(\varphi)  e^{i H t} =
e^{-i 2 \pi\varphi_2 I_z}  e^{-i H t} \left(e^{-i \frac{\pi \varphi_1}{2}
\sigma_2}\rho_0e^{i \frac{\pi \varphi_1}{2} \sigma_2}
\otimes \rho_0^C \otimes \rho_0^B  \right)   e^{i H t} e^{i 2 \pi\varphi_2 I_z}
\end{eqnarray}
After calculation of the trace with respect to 
the 
subsystems $A$ and $B$ of the density matrix (\ref{ev2}), we obtain $\rho^B(t)$ in the form  
\begin{eqnarray}
\rho^B(t)=e^{-i  \pi\varphi_2\sigma_3}  \tilde U(t) \Lambda(t) \tilde U^+(t) e^{i  \pi\varphi_2 \sigma_3},\;\;
\tilde U(t)= e^{-i \pi \tilde \beta(t) \sigma_3}
e^{-i \frac{\pi \beta_1(t)}{2} \sigma_2}e^{i \pi \tilde \beta(t) \sigma_3},
\end{eqnarray}
{{}which coincides with} form
(\ref{rhoBf},\ref{UB}) if we take 
$U(t)=e^{-i  \pi\varphi_2\sigma_3}  \tilde U(t)e^{i  \pi\varphi_2\sigma_3}$ {{}and
assume} $\beta_2(t) = \tilde \beta(t) + \varphi_2$. The later formula  is the  linear relation between 
$\beta_2$ and $\varphi_2$ mentioned in 
Sec.\ref{Section:3node}. 
Thus, varying the parameter $\varphi_2$, we can
obtain any required  value of the parameter $\beta_2$ at the needed time instant $t$. 
{{{}}This allows us to reduce  map (\ref{calM2nodes3}-\ref{betlam0}) to map 
(\ref{calM2nodes32}-\ref{betlam}).}

\subsection{Explicite form of the matrices $\gamma_i$ in eq.\ref{UrhoU}.}
\label{Section:appendixC}
Below we give  the list of matrices $\gamma_i$ representing the  basis of the Lie algebra of  $SU(4)$ \cite{GM}:
\begin{eqnarray}\label{gamma}
&&
\gamma_1=\left[
\begin{array}{cccc}
         0& 1& 0& 0\cr
         1& 0& 0& 0\cr
         0& 0& 0& 0\cr
         0& 0& 0& 0  
\end{array}
\right],\;\;
\gamma_2=\left[
\begin{array}{cccc}
         0&-i& 0& 0\cr
         i& 0& 0& 0\cr
         0& 0& 0& 0\cr
         0& 0& 0& 0  
\end{array}
\right],\;\;
\gamma_3=\left[
\begin{array}{cccc}
         1& 0& 0& 0\cr
         0&-1& 0& 0\cr
         0& 0& 0& 0\cr
         0& 0& 0& 0  
\end{array}
\right],\\\nonumber
&&
\gamma_4=\left[
\begin{array}{cccc}
         0& 0& 1& 0\cr
         0& 0& 0& 0\cr
         1& 0& 0& 0\cr
         0& 0& 0& 0  
\end{array}
\right],\;\;
\gamma_5=\left[
\begin{array}{cccc}
         0& 0&-i& 0\cr
         0& 0& 0& 0\cr
         i& 0& 0& 0\cr
         0& 0& 0& 0  
\end{array}
\right],\;\;
\gamma_6=\left[
\begin{array}{cccc}
         0& 0& 0& 0\cr
         0& 0& 1& 0\cr
         0& 1& 0& 0\cr
         0& \;\;0& 0& 0  
\end{array}
\right],\\\nonumber
&&
\gamma_7=\left[
\begin{array}{cccc}
         0& 0& 0& 0\cr
         0& 0&-i& 0\cr
         0& i& 0& 0\cr
         0& 0& 0& 0  
\end{array}
\right],\;\;
\gamma_8=\frac{1}{\sqrt{3}}\left[
\begin{array}{cccc}
         1& 0& 0& 0\cr
         0& 1& 0& 0\cr
         0& 0&-2& 0\cr
         0& 0& 0& 0  
\end{array}
\right],\;\;
\gamma_9=\left[
\begin{array}{cccc}
         0& 0& 0& 1\cr
         0& 0& 0& 0\cr
         0& 0& 0& 0\cr
         1& 0& 0& 0  
\end{array}
\right],\\\nonumber
&&
\gamma_{10}=\left[
\begin{array}{cccc}
         0& 0& 0&-i\cr
         0& 0& 0& 0\cr
         0& 0& 0& 0\cr
         i& 0& 0& 0  
\end{array}
\right],\;\;
\gamma_{11}=\left[
\begin{array}{cccc}
         0& 0& 0& 0\cr
         0& 0& 0& 1\cr
         0& 0& 0& 0\cr
         0& 1& 0& 0  
\end{array}
\right],\;\;
\gamma_{12}=\left[
\begin{array}{cccc}
         0& 0& 0& 0\cr
         0& 0& 0&-i\cr
         0& 0& 0& 0\cr
         0& i& 0& 0  
\end{array}
\right],\\\nonumber
&&
\gamma_{13}=\left[
\begin{array}{cccc}
         0& 0& 0& 0\cr
         0& 0& 0& 0\cr
         0& 0& 0& 1\cr
         0& 0& 1& 0  
\end{array}
\right],\;\;
\gamma_{14}=\left[
\begin{array}{cccc}
         0& 0& 0& 0\cr
         0& 0& 0& 0\cr
         0& 0& 0&-i\cr
         0& 0& i& 0  
\end{array}
\right],\;\;
\gamma_{15}=\frac{1}{\sqrt{6}}\left[
\begin{array}{cccc}
         1& 0& 0& 0\cr
         0& 1& 0& 0\cr
         0& 0& 1& 0\cr
         0& 0& 0&-3  
\end{array}
\right].
\end{eqnarray}
These matrices are used in 
the general expression for the unitary transformation $U^A$ in eq.(\ref{UrhoU}).

\newpage

{{{}}  
\subsection{Two-node sender: receiver's state distribution 
in the space of three parameters $(\lambda,\beta_1,\beta_2)$ }
\label{Section:appendixD}}

{{{}}
First of all, we show that the state distribution is uniform with respect to the parameter $\beta_2$. 
For this purpose, we fix some time instant $t_0$ (for definiteness, we take the same time instant 
as for all other experiments with the four-node chain, i.e., $t_0=6.4$; the reason for this choice 
is explained in the end of this subsection) and $\lambda^B=1$ (as an example) 
}, vary the parameters $\varphi_i$, $i=10,11,12$ 
and $\phi$ inside of the region (\ref{2varphi4nodes}) and calculate the corresponding values 
of the parameters 
$\lambda$ and $\beta_i$, $i=1,2$. As a  result, we obtain 
the creatable 
  states depending on three parameters $(\lambda,\beta_1,\beta_2)$ which 
  are represented by the black points 
{{{}}  (or black regions) in 
Fig.\ref{Fig:fournodes}. Therewith, the parameters $\beta_i$, $i=1,2$,
are depicted  along the ordinate axis  as a ''combined'' parameter
$z(\beta)=10 [10 \beta_1 ] +  [10\beta_2 ]$, where $[\cdot]$ means the integer part of a number. 
We see that the right boundary of the creatable region (shown in Figs.\ref{Fig:fournodes}a,c) 
is a step-like line, therewith 
 each step on this boundary corresponds to a fixed value of $[10\beta_1 ]$, while $[10\beta_2 ]$ varies
 within each particular step taking the values $0, 1, \dots, 9$.
The step-like  boundary  demonstrates the uniformity of 
the state distribution  with respect to 
the parameter 
$[10\beta_2 ]$ and, consequently, with respect to the parameter $\beta_2$ 
with the absolute accuracy $0.1$ (because the integer number $[10\beta_2 ]$ takes into account only the
first decimal of $\beta_2$). In the case of non-uniformity with respect to $[10\beta_2 ]$, 
the steps would be ''smoothed''.}
  Thus,  
the essential parameters are $\lambda$ and $\beta_1$   (similar to Sec.\ref{Section:3nodesnumerics}). 
This observation allows us to simplify  map (\ref{2M4nodes})  disregarding the parameter 
$\beta_2$ therein and thus passing to map (\ref{2M4nodessimpl}).

\begin{figure*}
 \hspace*{-2cm}  \epsfig{file=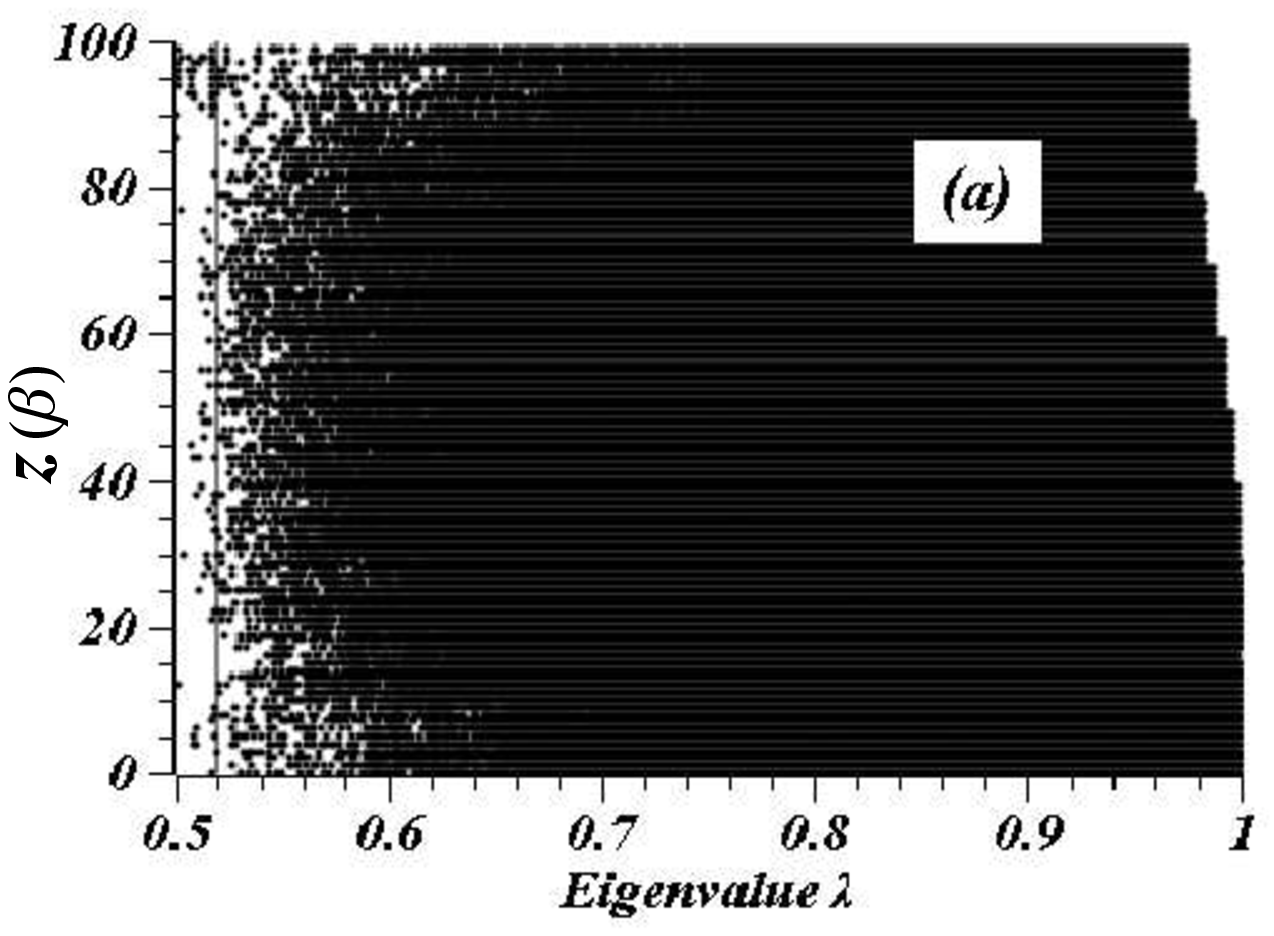,
  scale=0.44
   ,angle=0
}
\epsfig{file=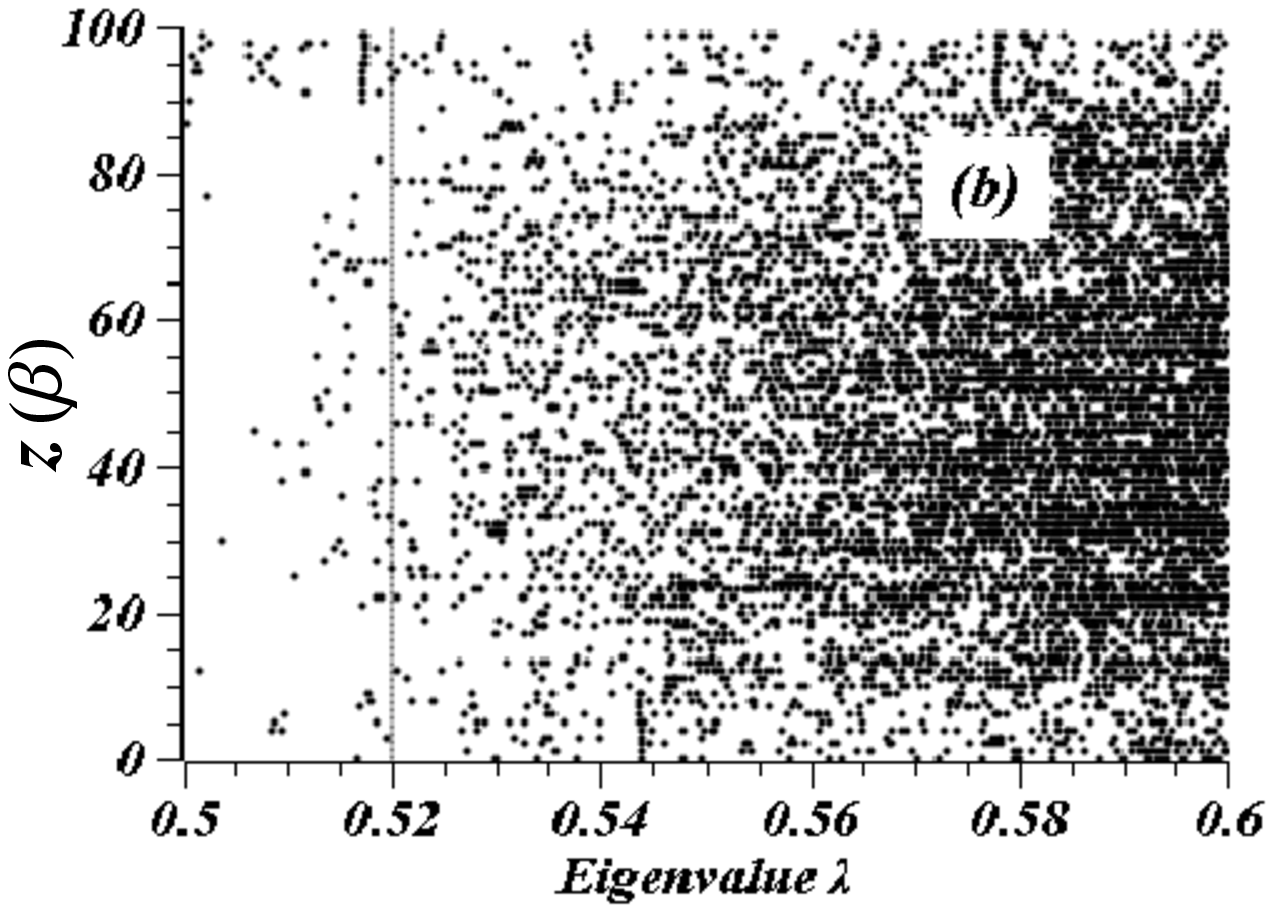,
  scale=0.44, 
   ,angle=0
}\epsfig{file=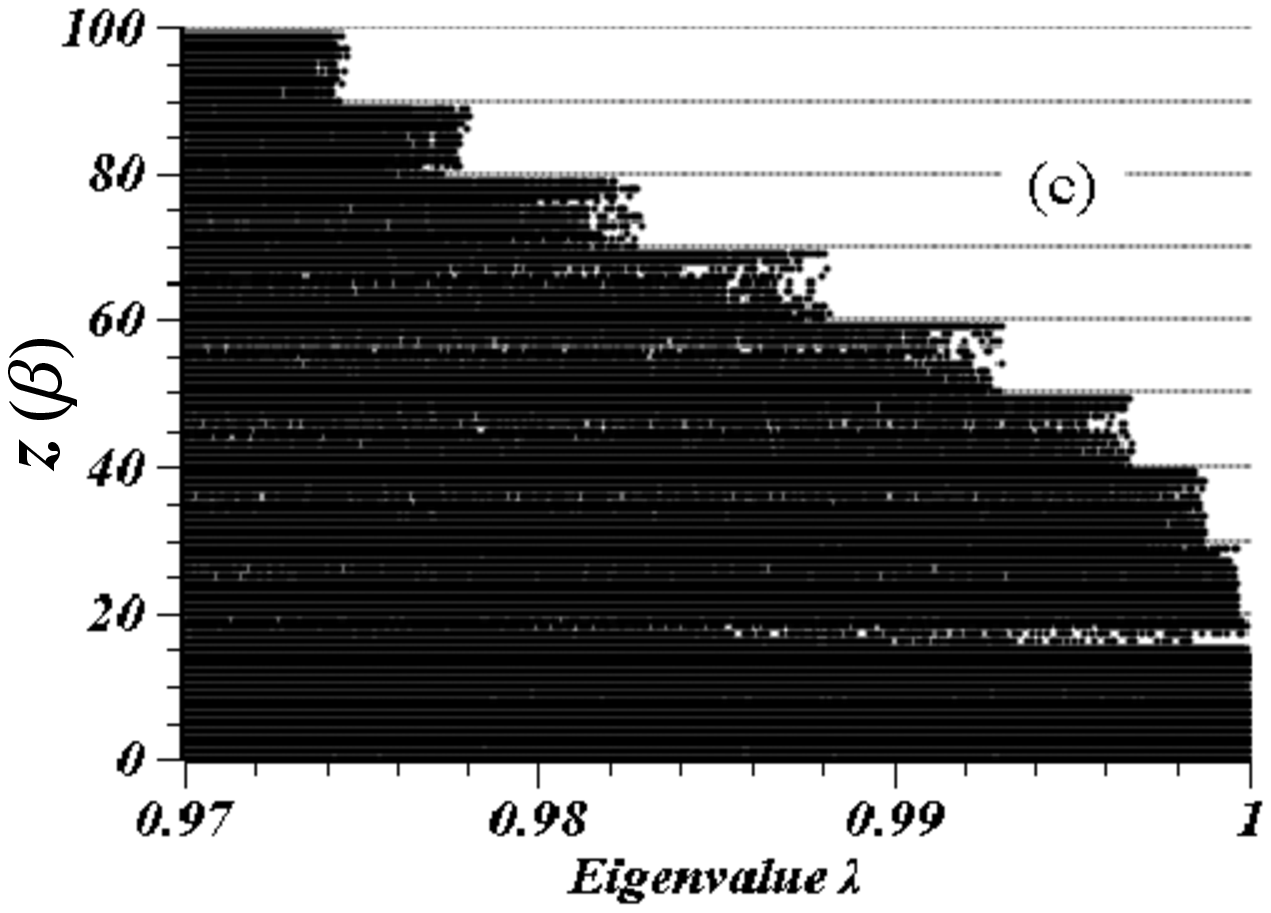,
  scale=0.44
   ,angle=0
}
\caption{
The three-parameter receiver state-space $(\lambda,\beta_1,\beta_2)$ (the black points)
of  map (\ref{2M4nodes}-\ref{2beta4nodes}) for the four node  chain with 
the two-qubit sender and the pure initial state ($\lambda^B=1$)  is considered 
at $t_0=6.4$.  
Here we combine the parameters $\beta_1$ and $\beta_2$ in the single parameter  
$z(\beta)=10 [10 \beta_1 ] +  [10 \beta_2]$. 
Apparently, the region of the available states is uniform  in $\beta_2$,
which is confirmed by  the  step-like behavior of the right boundary of the creatable region. 
(a) The  whole creatable region
of   the receiver's state-space. (b) The  hardly available area
of the creatable  states 
(the vertical stripe to the left from the dashed line).
(c) The right boundary of the  creatable region.
} 
  \label{Fig:fournodes} 
\end{figure*}

We see in  Fig.\ref{Fig:fournodes} that there is a region in the state-space
which may not be created by the local transformations of the subsystem $A$ (the right upper corner 
in Figs.\ref{Fig:fournodes}a,c). This is the unavailable region associated with  the initial state 
(\ref{inmatrices}) and $\lambda^B=1$. 
Other  initial states might have different unavailable regions.
There is another region in Fig.\ref{Fig:fournodes} which is hardly creatable. Conditionally,   this region 
may be taken as  a vertical stripe 
$\frac{1}{2}\le\lambda \le 0.52$, see the  left side from the dotted line in 
Figs.\ref{Fig:fournodes}a,b.
We shell give a following remark. Although we select the
parameters $\varphi_{10}$, $\varphi_{11}$ and $\varphi_{12}$
in map (\ref{2M4nodes}),
the similar   results (concerning the state transfer with $\lambda^B=1$)
are obtained for the other  selected triad of 
parameters: ($\varphi_{1}$, $\varphi_{2}$, $\varphi_{3}$), ($\varphi_{3}$, $\varphi_{4}$, $\varphi_{5}$),
($\varphi_{5}$, $\varphi_{6}$, $\varphi_{7}$), ($\varphi_{7}$, $\varphi_{8}$, $\varphi_{9}$),
($\varphi_{9}$, $\varphi_{10}$, $\varphi_{11}$). This fact supports our assumption that simplified 
map (\ref{2M4nodes}) produces  the maximal possible creatable region for the given choice of
 initial conditions.

{{{}}
Now, considering map (\ref{2M4nodessimpl}), 
we clarify our choice of the time instant $t_0=6.4$.  
It is reasonable to  select such a time instant for the state registration 
that corresponds to the maximal region creatable by  map  (\ref{2M4nodessimpl}). 
Generally speaking, this instant depends on the value of $\lambda^B$ in  initial state (\ref{inmatrices}).
However, to simplify the analysis, we use the same time instant $t_0$ for all values 
$\lambda^B$. Namely, let $t_0$ 
correspond to the maximal  creatable region obtained with $\lambda^B=1$ (a pure initial state). 
This value may be simply found from the numerical simulations of the creatable regions at 
different time instants inside of  the interval $[0, T\approx 10.2]$. 
As a  result, we obtain  $t_0\approx 6.4$.
}

\end{document}